\documentclass[apj,twocolumn,twocolappendix,numberedappendix]{openjournal}
\usepackage{newtxtext,newtxmath,enumitem,multirow,ctable}
\usepackage[T1]{fontenc}
\usepackage[breaklinks,colorlinks,citecolor=blue,urlcolor=blue]{hyperref}

\newcommand{\Alf}{{Alfv\'en}}
\newcommand{\bhat}{{\bf b}}

\newcommand{\etal}{et al.}
\newcommand{\orcidauthor}[3]{\author{\href{http://orcid.org/#1}{#2$^{#3}$}}}

\shorttitle{Non-Ideal MHD Limiters}
\shortauthors{Hopkins \etal}

\begin{document}
\title{\vspace{-0.8cm}Microphysical Regulation of Non-Ideal MHD in Weakly-Ionized Systems: \\ Does the Hall Effect Matter?\vspace{-1.5cm}}
\orcidauthor{0000-0003-3729-1684}{Philip F. Hopkins}{1,*}
\orcidauthor{0000-0001-8479-962X}{Jonathan Squire}{2}
\orcidauthor{0000-0003-2337-0277}{Raphael Skalidis}{1}
\orcidauthor{0000-0002-6810-1110}{Nadine H. Soliman}{1}
\affiliation{$^{1}$TAPIR, Mailcode 350-17, California Institute of Technology, Pasadena, CA 91125, USA}
\affiliation{$^{2}$Physics Department, University of Otago, 730 Cumberland St., Dunedin 9016, New Zealand}
\thanks{$^*$E-mail: \href{mailto:phopkins@caltech.edu}{phopkins@caltech.edu}},

\begin{abstract}
The magnetohydrodynamics (MHD) equations plus ``non-ideal'' (Ohmic, Hall, ambipolar) resistivities are widely used to model weakly-ionized astrophysical systems. We show that if gradients in the magnetic field become too steep, the implied charge drift speeds become much faster than microphysical signal speeds, invalidating the assumptions used to derive both the resistivities and MHD equations themselves. Generically this situation will excite microscale instabilities that suppress the drift and current. We show this could be relevant at low ionization fractions especially if Hall terms appear significant, external forces induce supersonic motions, or dust grains become a dominant charge carrier. Considering well-established treatments of super-thermal drifts in laboratory, terrestrial, and Solar plasmas as well as conduction and viscosity models, we generalize a simple prescription to rectify these issues, where the resistivities are multiplied by a correction factor that depends only on already-known macroscopic quantities. This is generalized for multi-species and weakly-ionized systems, and leaves the equations unchanged in the drift limits for which they are derived, but restores physical behavior (driving the system back towards slow drift by diffusing away small-scale gradients in the magnetic field) if the limits are violated. This has important consequences: restoring intuitive behaviors such as the system becoming hydrodynamic in the limit of zero ionization; suppressing magnetic structure on scales below a critical length which can comparable to circumstellar disk sizes; limiting the maximum magnetic amplification; and suppressing the effects of the Hall term in particular. This likely implies that the Hall term does not become dynamically important under most conditions of interest in these systems.
\end{abstract}
\keywords{magnetohydrodynamics (MHD) -- magnetic fields -- protoplanetary/circumstellar disks -- ISM: clouds -- star formation -- planet formation -- methods: numerical}
\maketitle

\section{Introduction}
\label{sec:intro}

Magnetohydrodynamics (MHD) is fundamental to almost all astrophysical systems. In particular, the last decade has seen a flurry of simulations and detailed models of MHD in weakly-ionized systems such as circumstellar/protoplanetary disks, disk and cool-star winds, or planet formation \citep{flock.2011:3d.nonideal.mhd.disk.sims,bai:2011.nonideal.effects.on.mri.accretion.in.planetary.disks,zhu:2014.non.ideal.mhd.vortex.traps,keith:2014.planetary.disk.ionization.model,tomida:2015.rad.nonideal.mhd.sf.sims,simon.2015:mri.driven.protoplanetary.disk.accretion,wurster.2016:non.ideal.mhd.braking,xu:2019.ppd.chemical.networks.for.nonideal.mhd,wang:2019.winds.nonideal.mhd,hennebelle:2020.nonideal.mhd.importance.star.formation,lee:2021.non.ideal.mhd.fx.planet.disks}. A large fraction of this literature adopts a common single-fluid ``non-ideal'' MHD equation for the neutrals plus charge carriers \citep{cowling:1976.mhd.book,ichimaru:1978.nonideal.mhd.deriv.and.assumptions,nakano:1986.nonideal.mhd.formulation.review}. 

These simulations have increasingly pushed to smaller scales and seen a variety of potentially important behaviors in many different regimes, emphasizing the importance of the different Ohmic, Hall, and ambipolar diffusion terms in systems with potentially strong shocks and gravitational and/or radiative forces \citep[for reviews see][]{teyssier:2019.numerical.mhd.review.powell.cant.work,wurster:2021.do.we.need.nonideal.mhd,tsukamoto:2022.nonideal.mhd.starformation.review}, as well as the importance of micron-sized dust grains as (sometimes dominant) charge carriers \citep{mestel:1956.charged.grains.magnetic.resistivity,elmegreen:1979.charged.grains.critical.magnetic.diffusion.coefficients,zhu:2014.dust.power.spectra.mri.turbulent.disks,tsukamoto:2022.importance.of.dust.sizes.for.nonideal.mhd,kawasaki:2022.dust.growth.ppd.nonideal.mhd.effects,kobayashi:2023.nonideal.disk.ion.vs.dust,marchand:2023.nonideal.mhd.with.grain.evolution}. But this can lead to some non-intuitive behaviors: while some properties become more ``hydrodynamic-like'' as the degree of ionization decreases ever smaller, others appear to show unique behaviors (especially in the Hall-dominated regime) which are qualitatively unlike hydrodynamics \citep{krasnopolsky:2011.sharp.hall.effects.in.disk.form.angmom,braiding.wardle:2012.hall.effect.star.form,tsukamoto:2015.hall.bimodal.sharp.change.disk.form,zhao:2020.hall.fx.ppds,wurster:2021.ambipolar.small.fx.starform.hall.fx.larger}. Moreover, as the ionization degree decreases, the implicitly assumed charge-carrier drift velocities can become extremely large, generating situations that are almost certainly highly unstable on unresolved plasma ``microscales'' or even unphysical \citep{mcbride:1972.two.stream.instability,mouschovias:1974.nonideal.mhd.ambipolar.fx.on.cloud.contraction,drake:1981.plasma.instabilities.high.drift,kulsrud:plasma.astro.book}. This can violate the foundational assumptions under which the usual non-ideal MHD expressions (both the equations themselves and the usual expressions for the various coefficients or currents) are derived. Such conditions will triggers micro-scale instabilities which induce strong ``anomalous resistivities'' far in excess of the ``classical'' resistivities usually adopted \citep{buneman:1958.instability.and.anomalous.resistivity,papadopoulos:1977.anomalous.resistivy.ionosphere,rowland:1982.anomalous.resistivity.aurora,galeev:1984.current.driven.instabilities.and.anomalous.resistivities,norman.heyvaerts:1985.anomolous.resistivity.current.instabilities.review,wahlund:1992.geo.applications.of.twostream.instability.observed.anomalous.resistivity,zweibel:2009.magnetic.reconnection.review}, but these have  largely been ignored in more recent astrophysical work \citep[though see][]{krasnopolsky:2010.still.need.anomolous.resistivity.for.disk.form,che:2017.anomalous.resistivity}. 

In this paper, we follow the usual derivation of the single-fluid non-ideal astrophysical MHD equations in order to derive the corresponding drift velocities, and highlight terms usually neglected that cannot be safely dropped in the limit of superthermal drift. We show how they can be approximately incorporated into modified non-ideal coefficients, and discuss the consequences of this, most notably in strongly suppressing gradients in the magnetic field below a critical scale $\ell_{\rm crit}$. This will limit the effects of the Hall term, likely causing it to lose its dynamical importance in most poorly-ionized plasmas of astrophysical interest. 

\section{Theoretical Background}
\label{sec:theory}

\subsection{Derivation}
\label{sec:deriv}

\begin{figure*}
	\centering\includegraphics[width=0.495\textwidth]{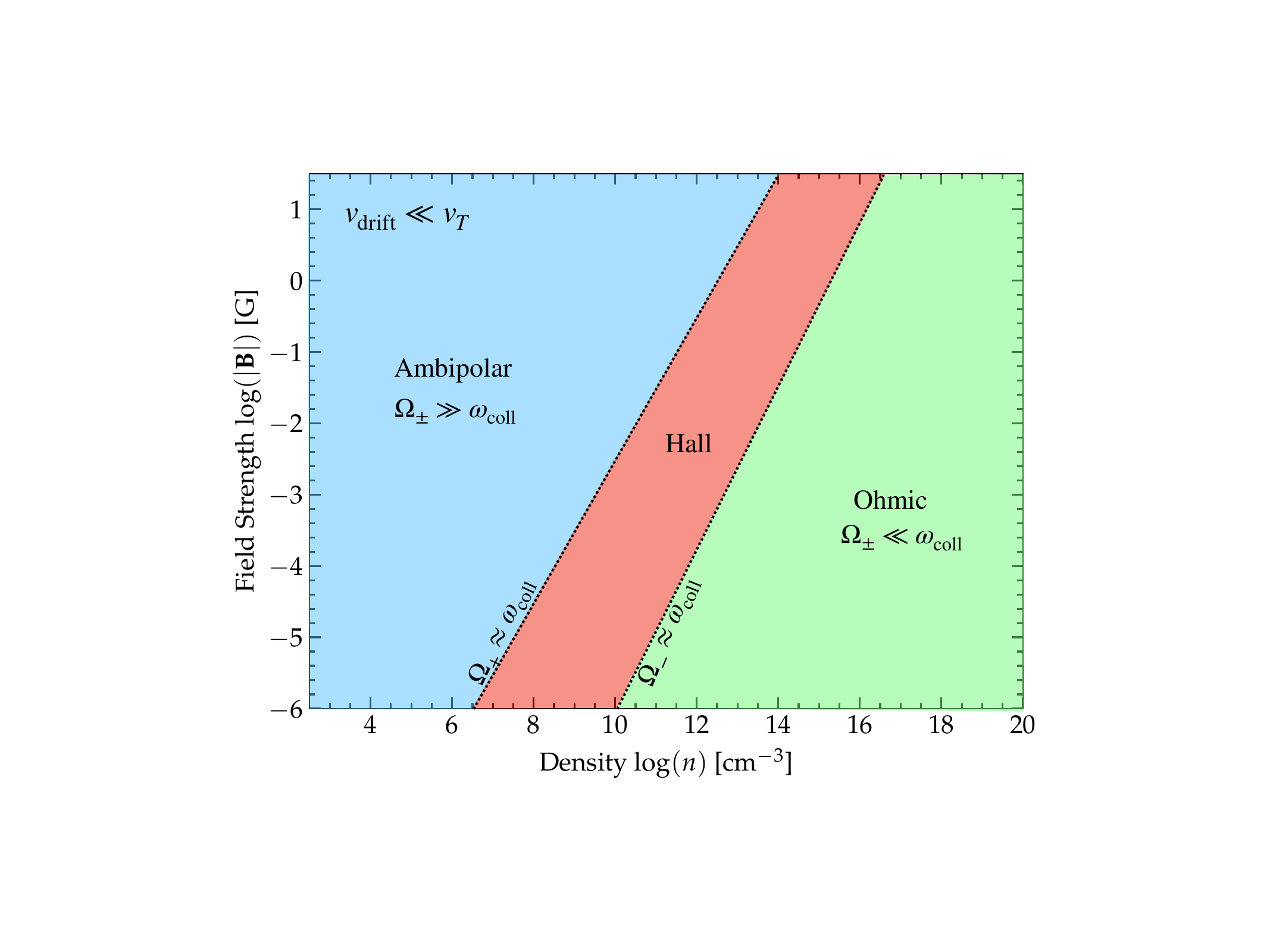} 
	\centering\includegraphics[width=0.495\textwidth]{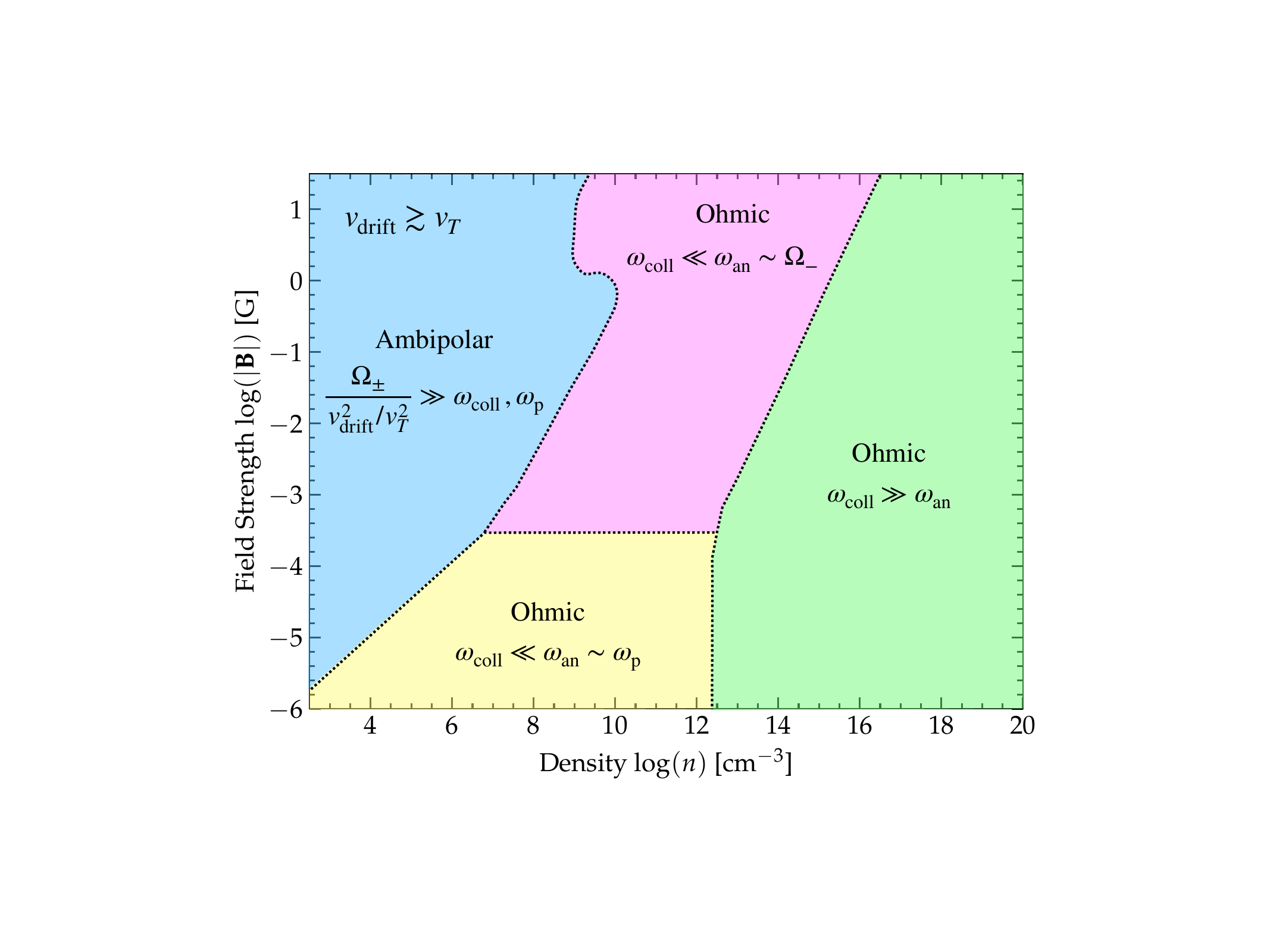} 
	\caption{Illustration of the regimes in density $\bar{n} = \bar{\rho}/m_{p}$ and magnetic field strength $\bar{B} \equiv | \bar{\bf B}|$, in a weakly-ionized gas ($n_{\pm} \ll n$), where different non-ideal MHD terms derived in \S~\ref{sec:deriv} (Ohmic resistivity $\eta_{O} {\bf J}$, Hall $\eta_{H} {\bf J} \times \bhat$, and ambipolar drift $\eta_{A} {\bf J} \times \bhat \times \bhat$) would each be relatively more important in the induction equation. Quantities like temperature, $v_{T}$, $\omega_{\rm coll}$, and ionization fractions at each $B$, $n$ are estimated as described in \S~\ref{sec:multispecies}-\ref{sec:plots}, for a multi-species (electrons, multiple ions, dust grains) system.
	{\em Left:} The case with ``classical'' non-ideal coefficients (\S~\ref{sec:deriv.nonideal.mhd}, Eq.~\ref{eqn:nonideal.classical}), which are valid only if the drift velocity between charge carriers $v_{\rm drift}\sim |{\bf J}|/ n_{-} |q_{-}| \sim c B/4\pi n_{-} |q_{-}| \ell_{B}$ ($\ell_{B} \sim |{\bf B}|/|\nabla {\bf B}|$) and ``slip'' velocity between neutrals and carriers (Eq.~\ref{eqn:vdrift}) are much less than the thermal velocities $v_{T}$ ($v_{\rm drift} \ll v_{T}$). This in turn requires that the magnetic gradient scale length $\ell_{B}$ is much larger than some $\ell_{\rm crit}$ (Eq.~\ref{eqn:lcrit}). 
	The multi-species coefficients come from NICIL (\S~\ref{sec:plots}) and follow the expressions in Appendix~\ref{sec:extra.multi}. 
	{\em Right:} The case if the implied drift becomes superthermal: $v_{\rm drift} \gtrsim v_{T}$. We use the corrected ``effective'' coefficients (\S~\ref{sec:superthermal}; Eqs.~\ref{eqn:etas.general}-\ref{eqn:etas.general.specific}), which account for superthermal drift and its effect both on direct particle collision rates and effective collisions via scattering (anomalous resistivity), assuming drift velocities are superthermal (here imposing $v_{\rm drift} = 10\,v_{T}$ for illustration), or $\ell_{B} \lesssim \ell_{\rm crit}$ (strong magnetic field gradients), with the effective $\eta_{O}^{\rm an}$ and $v_{T}$ for a multi-species dusty fluid following \S~\ref{sec:multispecies}, Eqs.~\ref{eqn:eta.an}-\ref{eqn:vT}.
	When the drift becomes superthermal, the effective Ohmic term becomes larger than the Hall term, and increases greatly in relative importance even at low densities. Where it predominates, we label whether the anomalous ($\omega_{\rm an}$)  or direct collision ($\omega_{\rm coll}$) term is most important.
	\label{fig:regimes}}
\end{figure*}

We begin with a derivation of the non-ideal MHD equations, following e.g.\ \citet{cowling:1976.mhd.book,ichimaru:1978.nonideal.mhd.deriv.and.assumptions,nakano:1986.nonideal.mhd.formulation.review}. This has been done before, but this helps motivate our discussion below, particularly regarding which terms can be dropped. Convenient variables used throughout this paper are defined in Table~\ref{tbl:variables}.

We follow the usual kinetic approach by computing moments of the distribution function of each species, assuming infinitesimally small averaging volumes (but implicitly larger than the interparticle separation and Debye length), with bulk velocities $\ll c$, quasi-neutrality, and collisional/exchange reactions being mass and momentum conserving (e.g.\ rest mass loss via photons is negligible). For an arbitrary set of species $j$ (which should formally include all isotopes/masses/ionization states/charges) this gives,
\begin{align}
\label{eqn:mom} & \frac{\partial (\rho_{j} {\bf u}_{j}) }{\partial t}  + \nabla \cdot \left( \rho_{j} \langle {\bf v}_{j}{\bf v}_{j} \rangle \right) =   \rho_{j} {\bf a}_{{\rm ext},\,j}  \\ 
\nonumber &\ \ +  \frac{q_{j} \rho_{j}}{m_{j}}\left[ {\bf E} + \frac{{\bf u}_{j}}{c} \times {\bf B} \right] + \sum_{i} \left[ {{\bf f}_{ji}} - \dot{\rho}_{ji}  {\bf u}_{j} + \dot{\rho}_{ij} {\bf u}_{i} \right],
\end{align}
in addition to the associated continuity equations $\partial_{t} \rho_{j} + \nabla \cdot (\rho_{j} {\bf u}_{j} ) = \sum_{i} \dot{\rho}_{ij} - \dot{\rho}_{ji}$. 
Here ${\bf u}_{j} = \langle {\bf v}_{j} \rangle_{\rm DF}$ is the local mean velocity of component $j$ averaged over its distribution function (DF); $\rho_{j}=m_{j}\,n_{j}$ is the mass density of species $j$ in terms of its particle mass $m_{j}$ and number density $n_{j}$; ${\bf a}_{{\rm ext},\,j}$ represents any external forces from e.g.\ gravity or radiation; ${\bf E}$ and ${\bf B}$ are the electric and magnetic fields; and ${\bf f}_{ji}$ and $\dot{\rho}_{ji}$ represent collisions and source/sink terms (via e.g.\ charge exchange, recombination, ionization) for each species. We can define these in terms of some effective rate coefficients: ${\bf f}_{ji} \equiv \rho_{j} \rho_{i} \gamma_{ji}\,({\bf u}_{i}-{\bf u}_{j})$ where $\rho_{i}\gamma_{ji} \equiv \rho_{i}\langle \sigma v \rangle^{ji}_{\rm DF} / (m_{j} + m_{i}) = \omega^{\rm coll}_{ji}$, and $\dot{\rho}_{ji} \equiv \rho_{j}\,\omega^{\rm ion/rec}_{ji}$. Note that $\omega^{\rm coll}$ (or $\gamma$) and $\omega^{\rm ion/rec}$ can in principle  be arbitrarily complicated functions of any other variables here including the various ${\bf u}_{j}$ (the exact dependence determined by the physics and particle types involved).

With the definition of total density  $\rho \equiv \sum_{j} \rho_{j}$ and mean fluid velocity ${\bf U} \equiv \rho^{-1} \sum_{j} \rho_{j} {\bf u}_{j}$, as well as current ${\bf J}  \equiv \sum_{j} q_{j} n_{j} {\bf u}_{j}$, we can sum over all species to give the familiar total momentum equation,
\begin{align}
\label{eqn:mom.tot} \frac{\partial (\rho {\bf U})}{\partial t} + \nabla \cdot \left( \rho{\bf U} {\bf U} \right) &= -\nabla \cdot \boldsymbol{\Pi} + \rho\,{\bf a}_{\rm ext} + \frac{{\bf J} \times {\bf B}}{c}, 
\end{align}
and continuity equation, $\partial_{t}\rho + \nabla \cdot (\rho {\bf U})= 0$, where $\rho {\bf a}_{\rm ext} \equiv \sum_{j} \rho_{j} {\bf a}_{{\rm ext},\,j}$, and $\boldsymbol{\Pi} \equiv \sum_{a} \boldsymbol{\Pi}_{a}$ with $\boldsymbol{\Pi}_{a} \equiv \rho_{a} \langle ({\bf v}_{a}-{\bf U})\,({\bf v}_{a}-{\bf U})\rangle$ is the total stress tensor. Note that the stress tensors are defined in the comoving frame of ${\bf U}$, which means $\boldsymbol{\Pi}_{a} = \rho_{a} \{ \delta{\bf u}_{a}  \delta{\bf u}_{a}   + \langle \delta {\bf v}_{a} \delta {\bf v}_{a} \rangle \}$ in terms of the drift velocity $\delta{\bf u}_{a} \equiv {\bf u}_{a}-{\bf U}$ (aka Reynolds stress or bulk flow pressure) and actual velocity dispersions $\delta {\bf v}_{a} \equiv {\bf v}_{a} - {\bf u}_{a}$. 

\begin{footnotesize}
\ctable[caption={{\small Variables Used Throughout\vspace{-0.2cm}}\label{tbl:variables}},center]{l l}{
}{
\hline
Name & Definition \\
\hline
${\bf E}$, ${\bf J}$, $c$ & electric field, current, speed-of-light \\
${\bf B}$, $B$, $\bhat$ & magnetic field ${\bf B} \equiv B\,\bhat$ \\
$q_{j}$, $m_{j}$, $\rho_{j}$ & charge, mass, density of species $j$ \\
$n_{j}$, $T_{j}$, ${\bf u}_{j}$ & number density, temperature, bulk velocity of species $j$ \\
\hline
$\omega_{ij}^{\rm coll}$, $\gamma_{ij}$ & collision rates \&\ rate coefficients between $i$ \&\ $j$ \\
$\tilde{\beta}_{j}$ & Hall or mobility parameter of species $j$ \\
$\xi_{j}$, $\Omega_{j}$ & mass fraction \&\ gyro frequency of species $j$ \\
$\rho$, ${\bf U}$ & total density and mean velocity (includes all $j$) \\
\hline
$\bar{X}$, $\delta {\bf X}$ & macroscopic average of ${\bf X}$, fluctuations $\delta {\bf X}$ \\ 
$\tilde{\bf J}_{A}$ & normalized Ampere's current $\tilde{\bf J}_{A} \equiv \nabla \times \bar{\bf B}$ \\
$\eta_{O,\,H,\,A}$ & Ohmic, Hall, ambipolar diffusivities \\
$\eta_{O,\,H,\,A}^{0}$ & diffusivities in the standard approximation \\
\hline
$\eta_{O}^{\rm an}$, $\omega^{\rm an}$ & diffusivity, effective collisions from anomalous resistivity \\
$\eta_{O,\,H,\,A}^{\rm eff}$ & approximate $\eta$ with anomalous \&\ finite-drift effects \\
${\bf v}_{\rm drift}$ & drift velocity between charge-carriers \\
${\bf v}^{\pm}_{\rm slip}$ & slip speed between neutrals and charged ($\pm$) species \\
\hline
$v_{T}$ & characteristic slow wavespeed (e.g.\ thermal speed) \\
$\ell_{B}$ & magnetic gradient scale-length \\
$\ell_{\rm crit}^{\rm drift}$, $\ell_{\rm crit}^{\rm slip}$ & value of $\ell_{B}$ below which $v_{\rm drift/slip}>v_{T}$ \\
\hline
}
\end{footnotesize}

Now consider a three-component system with a dominant set of neutrals, positive and negative charge carriers (labeled $n$, $+$, and $-$, respectively). We can repeat this derivation for an arbitrary set of species, and we do so in Appendix~\ref{sec:extra.multi} to validate our expressions. However in the general multi-species case many of our key results must be calculated numerically, while we can derive closed-form analytic expressions for the three-component case which inform our intuition and capture the salient dimensional scalings.

Define an ``ambipolar current'' ${\bf J}_{d} \equiv -(q_{-}\rho_{c}/m_{-})\,({\bf u}_{+} - {\bf u}_{n})$, where $\rho_{c} \equiv \rho_{+}+\rho_{-}$. We can then express the three velocities ${\bf u}_{n,+,-}$ in terms of ${\bf U}$, ${\bf J}$, and ${\bf J}_{d}$: ${\bf u}_{n} = {\bf U} - (m_{-}/q_{-}\rho)\,({\bf J}-{\bf J}_{d})$, ${\bf u}_{+}={\bf U} - (m_{-}/q_{-}\rho)\,[{\bf J} + (\rho_{n}/\rho_{c}) {\bf J}_{d}]$, ${\bf u}_{-} = {\bf U} - (m_{-}/q_{-}\rho)\,[-\{ (\rho_{+}+\rho_{n})/\rho_{-} \} {\bf J} + (\rho_{n}/\rho_{c}) {\bf J}_{d} ]$ \citep{cowling:1976.mhd.book,pinto.galli:2008.three.fluid.mhd.derivations}, and  rewrite our three Eq.~\ref{eqn:mom} for the evolution of ${\bf u}_{n,+,-}$ with an equivalent set for ${\bf U}$, ${\bf J}$, ${\bf J}_{d}$ via substitution. With Eq.~\ref{eqn:mom.tot} for ${\bf U}$, we have:
\begin{align}
\label{eqn:J} D_{t} & {\bf J} = \frac{|q_{-}|}{m_{-}} \nabla \cdot (\boldsymbol{\Pi}_{-} - \epsilon \boldsymbol{\Pi}_{+})  + |q_{-}| n_{-} \Omega_{1}  \left(\frac{c {\bf E}}{B} + {\bf U}\times \bhat \right)  \\
\nonumber & - {\bf G}_{1} -\Omega_{2} {\bf J} \times \bhat - \omega_{1} {\bf J}  +\Omega_{3} {\bf J}_{d} \times \bhat + \omega_{2} {\bf J}_{d}  \ , \\
\nonumber 
{\bf G}_{1} &\equiv  q_{+}n_{+}({\bf a}_{e,-}-{\bf a}_{e,+}) \ , \\ 
\nonumber 
\Omega_{1} &\equiv (1+\epsilon) \Omega_{-}
\ , \
\Omega_{2} \equiv \Omega_{-} [ 1 - {(1+\epsilon)\xi_{-}}] 
\ , \
 \Omega_{3} \equiv \xi_{n} \Omega_{+} \ , \\
\nonumber 
\omega_{1} &\equiv {(1+\epsilon) \omega^{\rm coll}_{-+} + \omega^{\rm coll}_{-n}} + \omega^{\rm rec}_{- n} 
\ , \
 \omega_{2}  \equiv \frac{\epsilon (\omega_{-n}^{\rm coll} - \omega_{+n}^{\rm coll})}{1+\epsilon} \ , 
\end{align}
where $D_{t} {\bf X} \equiv \partial_{t} {\bf X} + \nabla \cdot ({\bf U} {\bf X} + {\bf X} {\bf U} )$, $\bhat\equiv {\bf B}/|{\bf B}|$, $\epsilon \equiv (q_{+} m_{-})/(|q_{-}| m_{+})$, $\xi_{j} \equiv \rho_{j}/\rho$, $\Omega_{\pm} \equiv |q_{\pm}| B/(m_{\pm} c)$ are the gyro frequencies, 
and
\begin{align}
\label{eqn:Jd}  D_{t}& \left[ \xi_{n} ({\bf J}_{d} - {\bf J}) \right]  = - {\bf G}_{0}
+\Omega_{4} {\bf J} \times \bhat  + \omega_{3}{\bf J} - \omega_{4} {\bf J}_{d} \ , \\ 
\nonumber 
{\bf G}_{0} &\equiv \frac{|q_{-}|}{m_{-}} [{\xi_{n} \nabla \cdot \boldsymbol{\Pi} - \nabla \cdot \boldsymbol{\Pi}_{n} + \xi_{n} ({\bf a}_{e,-}\rho_{-} + {\bf a}_{e,+}\rho_{+} - {\bf a}_{e,n}\rho_{c}) }] \, ,\\
\nonumber 
 \Omega_{4} &\equiv  \xi_{n} \Omega_{-}  
\ , \ 
\omega_{3}  \equiv \omega_{-n}^{\rm coll} + \xi_{n}\left[(1+\epsilon)\,\omega_{n-}^{\rm ion} + \omega_{-n}^{\rm rec} \right] 
 \ , \\  
\nonumber 
\omega_{4}  & =   \frac{ \epsilon \omega^{\rm coll}_{-n} +  \omega^{\rm coll}_{+n}}{1+\epsilon} + {\xi_{n} \omega_{-n}^{\rm rec} + (1-\xi_{n}) \omega_{n-}^{\rm ion}}  \ .
\end{align}
Note these are similar to Eqs.~34 \&\ 35 in \citet{pinto.galli:2008.three.fluid.mhd.derivations}, except we have retained more general $\omega$, charge-exchange, and external acceleration terms. 

Eqs.~\ref{eqn:J}-\ref{eqn:Jd} can be rearranged to give an expression for ${\bf E}$: 
\begin{align}
\label{eqn:Ohms} {\bf E} &= - \frac{\bf U}{c}\times {\bf B}  + A_{0} {\Bigl [} \omega_{O} {\bf J} + \omega_{H} {\bf J} \times \bhat - \omega_{A} {\bf J} \times \bhat \times \bhat \\ 
\nonumber & \quad +  {{\bf D}_{t0}} + {\bf G}_{2} 
-\frac{|q_{-}|}{m_{-} } \nabla \cdot (\boldsymbol{\Pi}_{-} + \epsilon \boldsymbol{\Pi}_{+}) {\Bigr]} \ , \\
\nonumber {\bf D}_{t0} &\equiv D_{t}{\bf J} + \frac{\omega_{2}}{\omega_{4}} D_{t} [ {\xi_{n} ({\bf J}_{d}-{\bf J})}{} ] 
+ \frac{\Omega_{3}}{\omega_{4}}  D_{t} [ {\xi_{n} ({\bf J}_{d}-{\bf J})} ]  \times \bhat  \ , \\ 
\nonumber 
A_{0} &\equiv \frac{B}{|q_{-}| n_{-} c \, \Omega_{1}} \ , \ 
{\bf G}_{2} \equiv  {\bf G}_{1} + \frac{\omega_{2}}{\omega_{4}} {\bf G}_{0} + \frac{\Omega_{3}}{\omega_{4}} {\bf G}_{0} \times \bhat   \ , \\
\nonumber \omega_{O} &\equiv {\omega_{1}} -  \frac{\omega_{3} \omega_{2}}{\omega_{4}} \ , \ 
 \omega_{H} \equiv {\Omega_{2}} - \frac{\Omega_{4} \omega_{2}+\omega_{3} \Omega_{3}}{\omega_{4}} 
\ , \ 
 \omega_{A} \equiv \frac{\Omega_{4} \Omega_{3}}{\omega_{4}} \ .
\end{align}
Here ${\bf D}_{t0}$ collects the time dependence and advection terms and ${\bf G}_{2}$ collects the battery and diamagnetic terms.
This ${\bf E}$ can then be inserted into the induction equation $\partial_{t} {\bf B} = -c \nabla \times {\bf E}$.
It is convenient for this to define the variables:
\begin{align}
\label{eqn:adef} a_{i} &\equiv A_{0} \omega_{i} = \frac{B}{|q_{-}| n_{-} c} \frac{\omega_{i}}{\Omega_{1}} \ .
\end{align}
Note that $\omega_{O}$ is similar to an effective collision frequency $\omega^{\rm coll}$; $\omega_{H}$ is similar to the gyrofrequency $\sim \Omega_{-}$, and $\omega_{A}$ scales as $\sim \Omega_{-}\Omega_{+}/\omega^{\rm coll}$. 

Eq.~\ref{eqn:Ohms} (with $\partial_{t} {\bf B} = -c \nabla \times {\bf E}$) formally gives us an expression for the evolution of ${\bf B}$, but this is clearly not solveable exactly without kinetic methods that can resolve the plasma ``micro'' scales (e.g.\ gyro radii, collisional mean-free paths, etc.) and separately predict terms like the stress tensors $\boldsymbol{\Pi}_{j}$, time derivatives $D_{t}{\bf J}$, $D_{t}{\bf J}_{d}$, battery terms, etc. 
Even if one adopted some effective closure relations for these terms in Eqs.~\ref{eqn:mom}-\ref{eqn:Ohms} to attempt to explicitly integrate a ``three fluid'' method, the equations could, in many cases of interest, develop micro-scale (kinetic) or meso-scale (small-scale fluid) fluctuations in ${\bf B}$ or ${\bf J}$. As we are interested in the dynamics of large (``macro'') scales in the system, these could also contribute importantly to the large-scale dynamics. Thus, a crucial set of assumptions made in order to justify dropping certain terms in Eq.~\ref{eqn:Ohms} and therefore render it integrable on ``macro'' scales $\ell \sim \Delta x_{\rm macro}$, $T \sim \Delta t_{\rm macro}$ is to assume some scale separation of macro/meso/micro scales, and then to assume that fluctuations on the micro and meso-scales are negligible for the macro-scale quantities. 

Mathematically, one can represent this with some Reynolds expansion. Heuristically, define the large-scale average of some quantity ${\bf X}$ as $\bar{\bf X} \equiv \langle {\bf X} \rangle_{\mathcal{V}} = (\int_{\mathcal{V}} {\bf X} \,d^{3}{\bf x}\, dt) / (\int_{\mathcal{V}} d^{3}{\bf x}\, dt)$, integrating over some macro-scale hyperspace domain $\mathcal{V}$ which is large compared to micro/meso scales, but still small compared to global length scales of the problem. Then define ${\bf X} \equiv \bar{\bf X} + \delta{\bf X}$ where $\delta{\bf X}({\bf x},\,t)$ represents mesoscale fluctuations (and $\langle{\delta{\bf X}} \rangle_{\mathcal{V}}=0$). One can then simply insert these definitions to expand the induction equation $\partial_{t} {\bf B} = -c \nabla \times {\bf E}$ and obtain:
\begin{align}
\label{eqn:induction.full} \partial_{t} \bar{\bf B} &\approx -c \left[ \nabla \times \bar{\bf E} + \langle \nabla \times \delta {\bf E} \rangle_{\mathcal{V}} \right] \\ 
\nonumber &= \nabla \times \left[ 
\bar{\bf U} \times \bar{\bf B} + \langle \delta{\bf U} \times \delta {\bf B}\rangle_{\mathcal{V}} + 
\bar{a}_{O} \bar{\bf J} + \langle \delta a_{O} \delta {\bf J} \rangle_{\mathcal{V}} + ... \right] \\ 
\nonumber &+ \langle \nabla \times \left[  \delta {\bf U} \times \bar{\bf B} + \bar{\bf U} \times \delta {\bf B} + ...  \right] \rangle_{\mathcal{V}}
\end{align}
where the ``...'' represents the complete expansion of Eq.~\ref{eqn:Ohms} for ${\bf E}$, including all terms. Note that, as in large-scale dynamo theory \citep{rincon:2019.dynamo.theory.review}, this is not a series expansion, so we cannot prima facia assume the $\delta$ terms are small and drop them. Also of importance is that, given some state vector $\boldsymbol{\Psi} \equiv ({\bf U}, {\bf B}, {\bf J}, {\bf J}_{d}, ...)$, variables like $\omega_{i}$ or $a_{i}$ will not satisfy $\bar{a}_{i} = a_{i}(\bar{\boldsymbol{\Psi}})$, because they are non-linear functions of the state variables. Evaluating terms like $\langle \delta{\bf U} \times \delta {\bf B}\rangle_{\mathcal{V}}$ formally requires integrals over the entire spatial and time spectrum of fluctuations knowing the appropriate $N$-th order correlation functions between the fluctuations, so we stress that Eq.~\ref{eqn:induction.full} is no more integrable than Eq.~\ref{eqn:Ohms}, unless we impose additional assumptions. We introduce it for the sake of highlighting the Reynolds-type terms, $\langle \delta{\bf U} \times \delta {\bf B}\rangle_{\mathcal{V}}$, to show dimensionally how they enter the induction equation as these can behave like an anomalous resistivity, discussed below.

\subsection{The ``Standard'' Non-Ideal MHD Approximations}
\label{sec:deriv.nonideal.mhd}

To arrive at the standard formulation of one/two-fluid non-ideal MHD, we simplify Eq.~\ref{eqn:Ohms} or \ref{eqn:induction.full} by making the following assumptions \citep[see][]{ichimaru:1978.nonideal.mhd.deriv.and.assumptions,norman.heyvaerts:1985.anomolous.resistivity.current.instabilities.review,nakano:1986.nonideal.mhd.formulation.review,wardle.ng:1999.nonideal.coefficients,tassis.mouschovias:2005.nonideal.formulation.review.accretion,tassis.mouschovias:2007.nonideal.mhd.clouds.formulations.analytic.models,kunz:2009.formulation.nonideal.mhd}. 
{\bf (1)} Assume (given dimensional arguments since velocities are non-relativistic: $|{\bf U}|,\,c_{s},\,v_{\rm drift},\,v_{A} \ll c$) that displacement currents can be neglected so ${\bf J} \rightarrow {\bf J}_{A} \equiv (c/4\pi) \nabla \times {\bf B}$. 
{\bf (2)} Drop all terms from the second-line of Eq.~\ref{eqn:Ohms}. 
(a) Assume temporal/advection frequencies $D_{t}$ are much smaller than the collision+gyro frequencies $\omega_{i}$ ($|D_{t}| \ll \omega_{i}$), usually justified by assuming $|D_{t}| \sim 1/\Delta t_{\rm macro}$ varies only on ``macroscopic'' timescales because all currents come into local steady-state ($D_{t}\bar{\bf X}\rightarrow 0$) rapidly. 
(b) Assume that the battery terms are negligible, i.e.\ external accelerations are negligible compared to the typical non-ideal terms, $|{\bf a}_{\rm ext}| \ll |\omega_{i} v_{\rm drift}|$, ionization fractions are small, and the {\em difference} between external acceleration/pressure forces on neutrals and charge carriers is small.
(c) Assume the stresses are similarly negligible. Noting $\nabla \cdot (\boldsymbol{\Pi}_{-}+\epsilon \boldsymbol{\Pi}_{+}) \sim \rho_{-} \delta v^{2}/\lambda$ for some fluctuations on scale $\lambda$ (with velocity dispersion $\delta v$ including thermal+drift/bulk components), this is equivalent to assuming $|\delta v^{2}|/\lambda \ll |\omega_{i} v_{\rm drift}|$. This is usually justified by assuming the drift is sub-thermal, and the $\rho_{j} \langle ( {\bf v}_{j} - \langle {\bf v}_{j} \rangle) ( {\bf v}_{j} - \langle {\bf v}_{j} \rangle) \rangle$ component of the charge carrier pressure tensors come into thermal equilibrium (are Maxwellian and isotropic with a single temperature, $\delta v \sim v_{T}$) and vary only on macroscopic scales ($\lambda \sim \ell_{\rm macro}$).
{\bf (3)} In the various $\omega_{i}$ terms, assume collisions with any non-neutrals can be neglected ($\omega_{in} \gg \omega_{ij}$ for $j\ne n$), and ignore inelastic and charge-exchange reactions ($\omega^{\rm rec/ion} \ll \omega^{\rm coll}$). 
{\bf (4)} Assume the drift ($v_{\rm drift} \equiv {\bf u}_{+} - {\bf u}_{-}$) and slip ($v_{\rm slip}^{\pm} = {\bf u}_{\pm} - {\bf u}_{n}$) velocities are vanishingly small compared to the thermal and \Alf\ and other wave speeds of all species. This means (a) that the non-isotropic/Reynolds/bulk-flow components of the pressure tensors $\boldsymbol{\Pi}^{\rm aniso}_{j} = \rho_{j} ({\bf u}_{j} - {\bf U}) ({\bf u}_{j} - {\bf U})$ can be neglected, (b) that certain instabilities and anomalous resistivities can be neglected (discussed below), and (c) that cofficients like $\langle \sigma v \rangle_{\rm DF}^{ji}$ (equivalently $\gamma_{ji}$ or $\omega^{\rm coll}_{ji}$) can be taken to have their values for zero drift/slip, which (by definition) are independent of the drift speeds and therefore no longer complicated functions of ${\bf J}$, ${\bf J}_{d}$, ${\bf U}$, and their derivatives.
{\bf (5)} Assume that fluctuations on unresolved (micro/meso) scales are vanishingly small, so all ``$\delta{\bf X}$'' terms in Eq.~\ref{eqn:induction.full} can be neglected and all coefficients like $\omega_{i}$ are functions only of the mean macro-scale average values $\bar{\bf B}$, $\bar{\bf J}$, etc. 
Thus for all variables ${\bf X} \rightarrow \bar{\bf X}$.
Note that this assumption is often implied to be equivalent to the scale hierarchy assumption $\Delta x_{\rm macro} \gg \Delta x_{\rm micro}$, but from the discussion in \S~\ref{sec:deriv} it is obvious that is not sufficient.

With these assumptions, applied to Eq.~\ref{eqn:Ohms} to obtain ${\bf E}\approx \bar{\bf E}$ and $\partial_{t} \bar{\bf B} \approx -c \nabla \times \bar{\bf E}$ , we finally obtain a relatively simple equation for $\bar{\bf B}$: 
\begin{align}
\nonumber \partial_{t}\bar{\bf B} &\approx \nabla \times \left[   \bar{\bf U} \times \bar{\bf B}  
- \eta^{0}_{O}\,\tilde{\bf J}_{A} 
- \frac{\eta^{0}_{H}}{\bar{B}}\tilde{\bf J}_{A}\times \bar{\bf B}  
+ \frac{\eta^{0}_{A}}{\bar{B}^{2}} \tilde{\bf J}_{A} \times \bar{\bf B} \times \bar{\bf B} \right] \ , \\ 
\label{eqn:nonideal.classical} \tilde{\bf J}_{A} &\equiv \nabla \times \bar{\bf B} \ , \ \eta^{0}_{O}=\frac{c^{2} a_{O}^{0}}{4\pi}
 \ , \ \eta^{0}_{H}=\frac{c^{2} a_{H}^{0}}{4\pi}
  \ , \ \eta^{0}_{A}=\frac{c^{2} a_{A}^{0} }{4\pi}\ .
\end{align}
The $a^{0}_{i}$ notation denotes the value of $a_{i}(\bar{\boldsymbol{\Psi}})$ (Eq.~\ref{eqn:adef}) assuming all state variables have their macroscopic mean values, and dropping terms and simplifying by assuming negligible drift/slip speeds per assumptions {\bf (1)}-{\bf (5)}.

We can now define the drift speeds implied by this formulation in terms of the variables that are evolved in the fluid model:
\begin{align}
\label{eqn:vdrift} {\bf v}_{\rm drift} &\equiv \bar{\bf u}_{+} - \bar{\bf u}_{-} = \frac{\bar{\bf J}}{n_{-} |q_{-}|} = \frac{c}{4\pi \bar{n}_{-} |q_{-}|} \nabla \times \bar{\bf B} \ , \\ 
\nonumber {\bf v}_{\rm slip}^{+} &\equiv \bar{\bf u}_{+} - \bar{\bf u}_{n} = \frac{\epsilon}{(1+\epsilon)} \frac{\bar{\bf J}_{d}}{\bar{n}_{+} q_{+} } 
= \frac{\epsilon \omega^{0}_{-n} {\bf v}_{\rm drift}}{\epsilon \omega^{0}_{-n} + \omega^{0}_{+n}} + {\bf v}_{\rm slip,\bot}
\ ,
\\ 
\nonumber {\bf v}_{\rm slip}^{-} &\equiv \bar{\bf u}_{-} - \bar{\bf u}_{n} = {\bf v}_{\rm slip}^{+} - {\bf v}_{\rm drift}
=-\frac{\omega^{0}_{+n} {\bf v}_{\rm drift}}{\epsilon \omega^{0}_{-n} + \omega^{0}_{+n}}  + {\bf v}_{\rm slip,\bot}
\ ,
  \\ 
\nonumber {\bf v}_{\rm slip,\bot} &\equiv \frac{\bar{\rho}}{\bar{\rho}_{n}} \frac{\eta^{0}_{A}}{\bar{B}^{2}} (\nabla \times \bar{\bf B}) \times \bar{\bf B} \ .
\end{align}
The $\omega^{0}$ notation follows $a^{0}$, $\eta^{0}$, etc.
The ``effective'' maximum absolute value among the drift/slip speeds is given by: 
\begin{align}
v_{\rm drift,\,max} \approx  \sqrt{|{\bf v}_{\rm drift}|^{2} + |{\bf v}_{\rm slip,\bot}|^{2}} \ .
\end{align}

For dimensional analysis below, it is convenient to note that we can write the non-ideal terms as:
\begin{align}
\label{eqn:eta.units} \eta^{0}_{O}\,\tilde{\bf J}_{A}  &= \frac{\omega_{O}^{0}}{\Omega_{1}^{0}} {\bf v}_{\rm drift}\,  |\bar{\bf B}|  \sim \frac{\omega^{\rm coll}}{\Omega_{-}} v_{\rm drift} B, \\ 
\nonumber 
\frac{\eta^{0}_{H}}{\bar{B}}\tilde{\bf J}_{A}\times \bar{\bf B}  &=  \frac{\omega_{H}^{0}}{\Omega_{1}^{0}} {\bf v}_{\rm drift} \times \bar{\bf B} \sim \pm v_{\rm drift} B, \\
\nonumber 
\frac{\eta^{0}_{A}}{\bar{B}^{2}} \tilde{\bf J}_{A} \times \bar{\bf B} \times \bar{\bf B}  &=  \frac{\omega_{A}^{0}}{\Omega_{1}^{0}} ({\bf v}_{\rm drift} \times \bhat) \times \bar{\bf B} \sim \frac{\Omega_{+}}{\omega^{\rm coll}} v_{\rm drift} B,
\end{align}
where the latter dimensional scalings come from taking the mostly-neutral limit for $\omega_{O}^{0}/\Omega_{1}^{0} \approx \omega_{\rm coll}^{O,\,0}/\bar{\Omega}_{-}$ with $\omega_{\rm coll}^{O,\,0} = \omega_{+n}^{0} \omega_{-n}^{0} / \omega_{\rm coll}^{A,\,0}$ and $\omega_{\rm coll}^{A,\,0} = \omega_{+n}^{0}  + \epsilon  \omega_{-n}^{0} $; $\omega_{H}^{0}/\Omega_{1}^{0} \approx ( \omega_{+n}^{0}  - \epsilon  \omega_{-n}^{0} ) / \omega_{\rm coll}^{A,\,0} \sim \pm1$; and $\omega_{A}^{0}/\Omega_{1}^{0} \approx \bar{\Omega}_{+} / \omega_{\rm coll}^{A,\,0}$.

\subsection{The Superthermal Drift Limit}
\label{sec:superthermal}

Per \S~\ref{sec:deriv.nonideal.mhd}, the assumptions used in deriving the standard non-ideal formulation explicitly require 
\begin{align}
\label{eqn:consistency} {\rm MAX}(v_{\rm drift},\,v_{\rm slip}^{\pm}) \approx v_{\rm drift,\,max} \ll v_{T} \ ,
\end{align}
where $v_{T}$ is some characteristic slow wavespeed of the different components. 
Specifically, this is explicitly part of assumption {\bf (4)} (\S~\ref{sec:superthermal:collisionrates}; see also Appendix~\ref{sec:extra.multi}), and implicitly (for reasons discussed in more detail below) part of assumptions {\bf (2)} \&\ {\bf (5)} (assumptions {\bf (1)} and {\bf (3)}, on the other hard, are not so directly affected by violations of Eq.~\ref{eqn:consistency}, though we discuss what might happen if they break down below as well). 
From Eq.~\ref{eqn:vdrift} for $v_{\rm drift}$ and $v_{\rm slip}^{\pm}$, we see that 
Eq.~\ref{eqn:consistency} is equivalent to: 
\begin{align}
\label{eqn:lcrit} \ell_{B} &\equiv \frac{\bar{B}}{|\nabla \times \bar{\bf B}|} \gg \ell_{\rm crit} \equiv {\rm MAX}\left[ \ell_{\rm crit}^{\rm drift} \ , \ \ell_{\rm crit}^{\rm slip} \right] \ , \\ 
\nonumber & \ell_{\rm crit}^{\rm drift} \equiv \frac{\bar{B}\,c}{4\pi \bar{n}_{-} |q_{-}| v_{T}} \ , \ 
\ell_{\rm crit}^{\rm slip} \equiv \frac{\eta_{A}^{0}}{\bar{\xi}_{n} v_{T}} =  \frac{\bar{B}^{2} \bar{\xi}_{n}}{4\pi \alpha_{n}^{0} v_{T}} \ , 
\end{align}
where $\alpha_{n}^{0} \equiv \bar{\rho}_{c} \omega_{4}^{0} = \rho_{-}\omega_{-n}^{0} + \rho_{+}\omega_{+n}^{0}$. We will show below that nothing in the standard formulation of Eq.~\ref{eqn:nonideal.classical} prevents this condition from being violated. So what will happen if superthermal drifts arise ($\ell_{B} \lesssim \ell_{\rm crit}$)?

\subsubsection{Modified Collision Rates}
\label{sec:superthermal:collisionrates}

Superthermal drift/slip speeds self-evidently violate assumption {\bf (4)} in \S~\ref{sec:deriv.nonideal.mhd}. This means that the collision rates usually adopted, which assume negligible bulk relative drift/slip velocities of different species (so they depend on thermal velocities but not the various $\Delta u_{ij} \equiv |{\bf u}_{i} - {\bf u}_{j}|$): $\omega_{ij}^{0} \equiv \omega_{ij}(\Delta u_{ij} \rightarrow 0)$, are simply incorrect, even if all the other assumptions in \S~\ref{sec:deriv.nonideal.mhd} held perfectly. 

If we retain all other assumptions from \S~\ref{sec:deriv.nonideal.mhd}, but allow the various $\omega_{ij}$ to be functions of $\Delta u_{ij}$ as above, then Eqs.~\ref{eqn:nonideal.classical} \&\ \ref{eqn:vdrift} apply but with modified coefficients $\eta \ne \eta^{0}$. 
In detail, the full collisional and inelastic couplings $\omega_{ij}$ are complicated non-linear (even non-monotonic) functions of the different relative velocities $\Delta u_{ij}$, which means they become implicit functions of ${\bf U}$, ${\bf J}$, ${\bf J}_{d}$, \citep{elmegreen:1979.charged.grains.critical.magnetic.diffusion.coefficients,draine.salpeter:ism.dust.dynamics,pandey.wardle:2006.nonideal.coefficients.and.dynamics,pinto.galli:2008.momentum.transfer.coefficients.for.weakly.ionized.systems,oberg:ppd.disk.chemistry.review}. This means the coefficients $\eta$ must be calculated numerically, which we show for simple three and five-species models in \S~\ref{sec:extra.multi}. However, in \S~\ref{sec:epstein.deriv} we use our three-species derivation (\S~\ref{sec:deriv}) to show that the leading-order behavior in the superthermal limit can be approximated with simple scalings. Briefly, in the limit $\Delta u_{ij} \gg v_{T,i},\,v_{T,j}$, then $\omega_{ij} \propto \langle \sigma v_{ij} \rangle \propto \Delta u_{ij}$ increasingly moves towards the Epstein limit in which the drag force scales with the interparticle velocity. Thus, the collision rates scale with the drift/slip speed $\omega_{ij} \sim n_{j} \langle \sigma v_{ij} \rangle^{0} \sqrt{1 + \Delta u^{2}_{ij}/v_{T,\,ij}^{2}}$ where $\Delta u_{ij} \sim v_{{\rm drift/slip},\,ij}$. If we simply multiply the collision rates by some factor $f_{\omega}$, we see from Eq.~\ref{eqn:Ohms} that $\eta_{O} \propto \omega_{O}$ scales $\propto f_{\omega}$, while $\eta_{H}$ is independent of $f_{\omega}$ and $\eta_{A} \propto \omega_{A}^{-1} \propto f_{\omega}^{-1}$. Thus after consideration of which species and relative speeds $\Delta u_{ij}$ appear in each, we obtain the approximate Eq.~\ref{eqn:epstein.scalings}: $\eta_{O} \sim \eta_{O}^{0}\,[1 + (|v_{\rm slip,\,eff}|^{2} + |{\bf v}_{\rm drift}^{0}|^{2})/v_{T,\,\rm eff}^{2}]^{1/2}$, $\eta_{H} \sim \eta_{H}^{0}$, $\eta_{A} \sim \eta_{A}^{0}\,[1 + |v_{\rm slip,\,eff}|^{2} /v_{T,\,\rm eff}^{2}]^{1/2}$, where $v_{\rm slip,\,eff}$ is defined there.

\begin{figure}
	\centering\includegraphics[width=0.95\columnwidth]{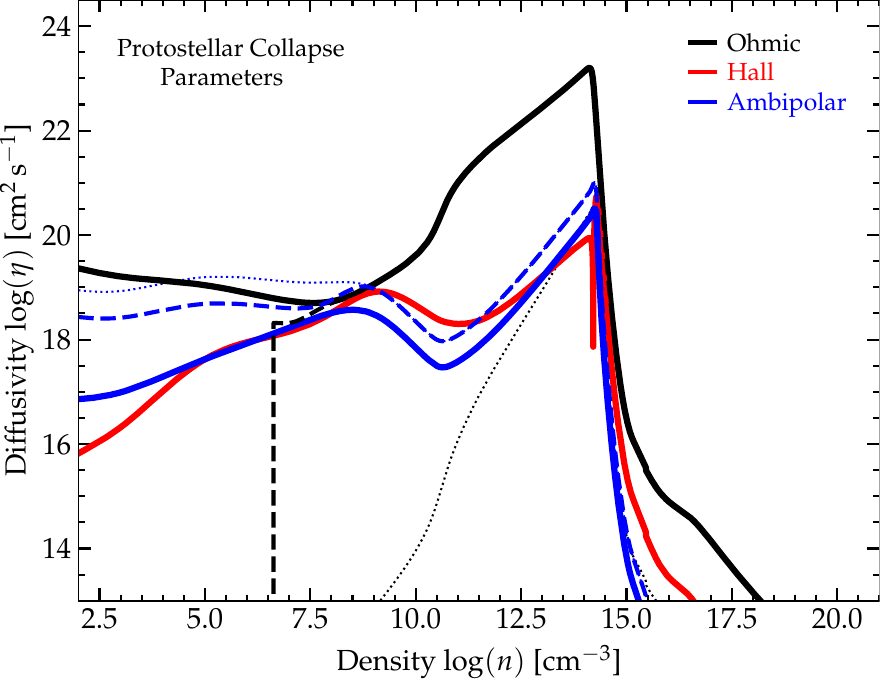} 
	\centering\includegraphics[width=0.95\columnwidth]{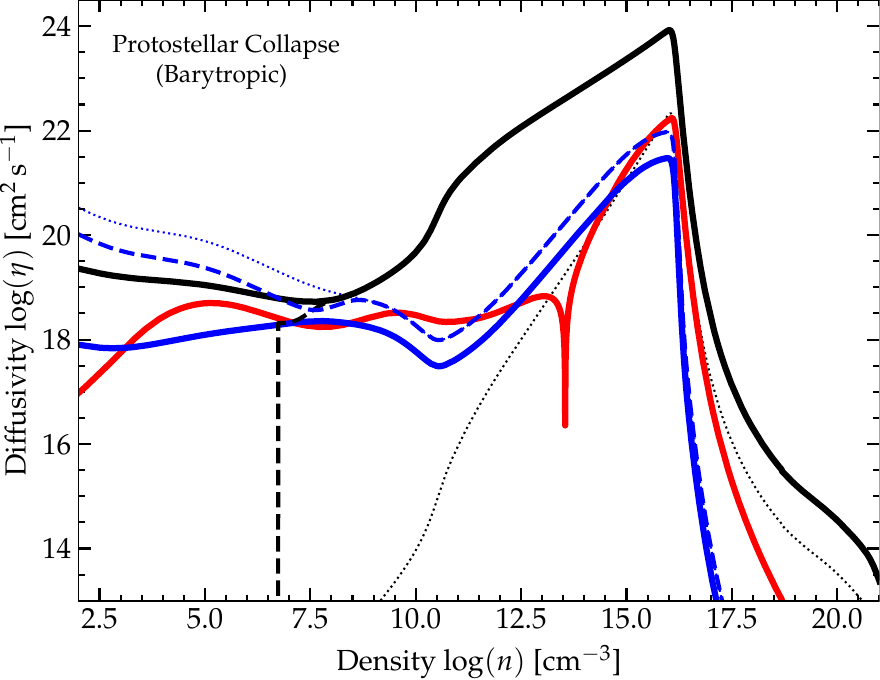} 
	\centering\includegraphics[width=0.95\columnwidth]{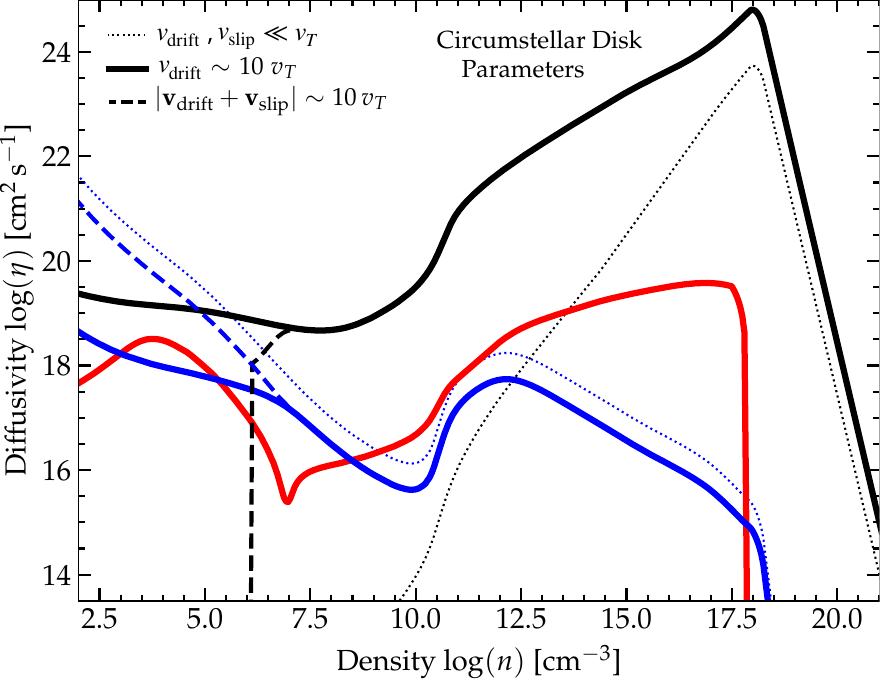} 
	\caption{Ohmic, Hall, and ambipolar diffusivities $\eta_{O,H,A}$ computed for a multispecies, weakly-ionized gas with total (mostly neutral) density $n$.	We compare values for slow drift+slip $v_{\rm drift},\,v_{\rm slip} \ll v_{T}$ ($\ell_{B} \gg \ell_{\rm crit}$; {\em dotted}) vs superthermal drift ($v_{\rm drift}\sim 10\,v_{T}$, $\ell_{B} \sim 0.1\,\ell_{\rm crit}^{\rm drift}$; {\em solid}) including the proposed corrections in Eq.~\ref{eqn:etas.general.specific} to account for enhanced scattering. We also compare $\ell_{B}\sim0.1\,\ell_{\rm crit}$ ($v_{\rm drift,\,max}=|{\bf v}_{\rm drift}+{\bf v}_{\rm slip}| \sim 10\,v_{T}$; {\em dashed}), which has superthermal slip but not drift at low-$n$. 
	For Hall, the lines overlap. 
	We consider three models for $B$, $T$, ionization, etc. from NICIL (\S~\ref{sec:plots}): a protostellar collapse model with non-ideal MHD ({\em top}), a barytropic equation-of-state model with stronger $B$ ({\em middle}), and a circumstellar disk midplane model with lower $T$ and weaker ionization ({\em bottom}). 
	The models change the value of $\eta_{O,H,A}$ (e.g.\ collapse models become hot at $n \gtrsim 10^{14} {\rm cm^{-3}}$, causing rapid ionization and decreasing $\eta$), but do not change the systematic offsets of interest here. 
	With $\ell_{B} \gg \ell_{\rm crit}$ (sub-thermal drift), the system transitions from ambipolar to Hall to Ohmic with increasing $n$. Superthermal drift/slip greatly enhances $\eta_{O}$ and eliminates the Hall regime.
	\label{fig:etas}}
\end{figure}

\subsubsection{Anomalous Resistivity: Dropped Terms and Instabilities}
\label{sec:superthermal:instabilities}

Superthermal drifts also imply that many of the dropped terms associated with assumptions {\bf (2)}, {\bf (4)} \&\ {\bf (5)} in \S~\ref{sec:deriv.nonideal.mhd} can no longer be dropped. It has been well-known for decades that superthermal drifts are generically unstable on micro and mesoscales \citep[see][and references therein]{kulsrud:plasma.astro.book}. For systems with non-trivial chemistry (many charged species), there are a potentially infinite number of such instabilities, some relatively well-studied such as the dozen distinct current-driven instabilities (e.g.\ two-stream, ion-acoustic, ion-cyclotron, whistler, Buneman) reviewed in \citet{kindel:1971.current.driven.instabilities,krall:1971.lower.hybrid.drift.instabilities,norman.smith:1978.instabilities.kinetic.scale.summary.for.proton.electron.plasma,norman.heyvaerts:1985.anomolous.resistivity.current.instabilities.review} or tearing \citep{mestel:1968.magnetic.breaking} or hybrid-mode \citep{davidson:1975.lower.hybrid.drift.instability.experiment.and.analytic.derivs} instabilities; others that have only barely been explored, such as the various novel instabilities of mixed neutral-charged systems identified in \citet{huba:1991.magnetic.drift.instabilities,kamaya:2000.alfven.thermal.twofluid.instabilities.poorly.ionized.gas,mamun:2001.instabilities.partially.ionized.long.wavelength,nekrasov:2008.plasma.instab.strongly.coupled.regimes,nekrasov:2008.compressible.streaming.instabilities.multicomponent}, or the families of intrinsically two-fluid (neutrals plus ions+electrons) drift instabilities in \citet{tytarenko:two.fluid.drift.instabilities,nekrasov:2009.plasma.instab.strongly.coupled.regimes} or the charged dust+gas ``resonant drag instabilities'' in \citet{squire.hopkins:RDI,hopkins:2018.mhd.rdi,seligman:2018.mhd.rdi.sims,hopkins:2019.mhd.rdi.periodic.box.sims}. These instabilities grow rapidly on micro/meso-scales, producing nonlinear fluctuations that violate assumptions {\bf (2)}, {\bf (4)} \&\ {\bf (5)}. 

Consider a simple example. An explicit assumption in deriving the standard MHD equations is that the pressure tensors $\boldsymbol{\Pi}_{j} = \rho_{a}({\bf u}_{a}-{\bf U})({\bf u}_{a}-{\bf U}) + \rho_{a} \langle \delta {\bf v}_{a} \delta {\bf v}_{a} \rangle$ are dominated by an isotropic component ($\langle \delta {\bf v}_{a} \delta {\bf v}_{a} \rangle \approx \mathbb{I}\, v_{T,a}^{2}$), i.e.\ that the Reynolds/bulk component $({\bf u}_{a}-{\bf U})({\bf u}_{a}-{\bf U}) \sim \mathcal{O}(v_{\rm drift}^{2})$ of the tensor can be neglected. If  $v_{{\rm drift},a} \gtrsim v_{T,a}$ this is self-evidently invalid. Moreover if the instabilities can source non-linear fluctuations in ${\bf v}_{\rm drift}$ on some scale $\ell \ll \ell_{B}$, then the anisotropic pressure term in Eq.~\ref{eqn:Ohms}, $\langle \nabla \cdot \boldsymbol{\Pi}_{-} \rangle_{\mathcal{V}} \sim \langle \rho_{-} \delta{\bf v}_{\rm drift} \cdot \nabla \delta {\bf v}_{\rm drift} \rangle_{\mathcal{V}}$, will generically give rise to a non-vanishing term of order $|q_{-}| n_{-}\,|v_{\rm drift}|^{2}/\ell$ along the local direction of $\hat{\bf v}_{\rm drift}$ and therefore $\hat{\bf J}$. This will act like an ``effective'' resistivity. Comparison of the magnitude of this term to the ``classical'' resistivity term $\omega_{O} {\bf J} \sim \omega_{\rm coll} |q_{-}| n_{-} v_{\rm drift} $ in Eq.~\ref{eqn:Ohms} shows their ratio is $\sim v_{\rm drift} / (\ell\,\omega_{\rm coll})$. Thus if we associate a characteristic frequency $\omega_{\delta} \sim v_{T}/\ell$ to the fluctuations, we see that this specific term will contribute a resistivity $\sim \omega_{\delta}\,(v_{\rm drift}/v_{T})$ which is larger than the classical resistivity if this is larger than the collision frequency. We can straightforwardly obtain similar results for terms like $\langle \delta {\bf u}_{-} \times \delta {\bf B} \rangle$, which for non-linear fluctuations could give rise to resistivities as large as the gyro or plasma frequencies \citep{sagdeev:1966.nonlinear.plasma.resistivities.etc,frieman:1982.nonlinear.perturb.mhd.eqns,yoon:2006.quasi.linear.anomalous.resistivity.theory,graham:2022.direct.obs.anomalous.resistivity.ion.cyclotronl.timescale}.

These effects are ``anomalous resistivity,'' which is well-studied in laboratory fusion plasmas, ionospheric dynamics, and Solar physics \citep{buneman:1958.instability.and.anomalous.resistivity,papadopoulos:1977.anomalous.resistivy.ionosphere,rowland:1982.anomalous.resistivity.aurora,galeev:1984.current.driven.instabilities.and.anomalous.resistivities,wahlund:1992.geo.applications.of.twostream.instability.observed.anomalous.resistivity}. In these applications, it has been shown that when $v_{\rm drift}$ exceeds speeds like the ion thermal speed and/or \Alf\ speed, the measured resistivity rises rapidly (often by orders of magnitude as the drift speed increases by a small amount, from e.g.\ just below the thermal speed to $\sim 1.5$ times larger than the thermal speed; see \citealt{davidson:1975.lower.hybrid.drift.instability.experiment.and.analytic.derivs,gentle:1978.resistivity.observations.in.lab.plasmas.scalings,petkaki:2003.anomalous.resistivity.nonmaxwellian.terms.rapid.rise,graham:2022.direct.obs.anomalous.resistivity.ion.cyclotronl.timescale}). Considerable theoretical and experimental effort has gone into characterizing and modeling it in these fields \citep[see references above and][]{dobrowlowny:1974.theory.of.anomalous.resistivity.in.turbulent.plasma,ugai:1984.anomalous.resistivity.fast.reconnection,bychenkov:1988.ion.acoustic.turbulence.sims.vs.analytic.theory,uzdensky:2003.anomalous.resistivity.derivation.and.solar.applications,yoon:2006.quasi.linear.anomalous.resistivity.theory,lee:2007.current.driven.corona.theory.for.anomalous.resistivity,roytershteyn:2012.lower.hybrid.drift.instability.and.anomalous.resistivity,beving:2023.pic.simulations.anomalous.resistivity.up.to.plasma.frequency}, but it has largely been neglected in the applications in \S~\ref{sec:intro} (with some exceptions noted there). Detailed predictions in the superthermal regime clearly require explicit kinetic treatments (e.g.\ PIC simulations), and many of the non-linear outcomes remain poorly understood (particularly so for e.g.\ circumstellar disks as opposed to fusion plasmas). However, a generic result of these calculations and experiments is that the leading-order terms that arise act to oppose the drift (i.e.\ they act like resistivity), as the energy dissipated in fluctuations must ultimately come from the driving force (the drift current itself). Further, their ``effective'' resistivity scales with $\sim \omega_{\delta} \sim \omega_{\rm max}\,W$, where $\omega_{\rm max}$ is some fastest rate on which the modes can act (usually the gyro or plasma frequency, or some combination of the two; \citealt{davidson:1975.lower.hybrid.drift.instability.experiment.and.analytic.derivs,nayar:1978.resistivity.geomagnetic.tail.approaching.electron.gyro,norman.heyvaerts:1985.anomolous.resistivity.current.instabilities.review,treumann:2001.anomalous.resistivity.mini.review,graham:2022.direct.obs.anomalous.resistivity.ion.cyclotronl.timescale})\footnote{Note that much of the historical study of anomalous resistivity, as a result of its intended applications in laboratory and Solar plasmas, has focused on the high-density regime where $\omega_{pe} \gg \Omega_{e}$. Thus it is common to see $\omega_{\rm max} \sim \omega_{pe} (\delta v/v_{T})^{2}$ as a characteristic estimate. For systems with $\omega_{pj} \ll \Omega_{j}$ where collective effects are weak, the situation is more analogous to pitch-angle scattering where one generically obtains $\omega_{\delta} \sim \Omega_{j}\,(\delta v/v_{\rm eff})^{2}$ (with excitation above some ``effective'' thermal or \Alf\ speed if the heavier carriers and neutrals are well-coupled; \citealt{jokipii:1966.cr.propagation.random.bfield,kulsrud.1969:streaming.instability,coppi:1971.plasma.resistivity.low.Efield,schlickeiser:89.cr.transport.scattering.eqns}).} and $W \sim \langle \delta v^{2} \rangle / v_{0}^{2}$ is defined from some ``characteristic'' velocity $v_{0}$ above which the modes can grow (e.g.\ the thermal-ion or ideal-\Alf\ speed, depending on the mode). Generically, such instabilities saturate with $\omega_{\delta} \sim \omega_{\rm max}$ (see references above).\footnote{Consider the following heuristic ``derivation.'' Many of the salient instabilities have dispersion relations for which the growth rate can be (dimensionally) expressed as $\Gamma \sim \omega_{\rm max}\,(v_{\rm drift}/v_{\rm crit}-1)$ above some critical drift velocity $v_{\rm crit} \sim v_{T}$. In the weakly-ionized case of interest, the dominant damping mechanism is collisions, with frequency $\omega_{\rm coll}$. If $\omega_{\rm coll} \gg \omega_{\rm max}$, the instabilities will be suppressed or damped but in this case, one would (by definition) already be well into the ``classical'' Ohmic limit, and the maximum anomalous resistivity correction ($\propto (1+ \omega_{\rm max}/\omega_{\rm coll})$) would be small, so this has no effect on our results. But if $\omega_{\rm coll} \ll \omega_{\rm max}$, damping cannot compete with growth. Since the frequency $\omega_{\rm max}$ is much larger than macroscopic frequencies $\sim |{\bf u}|/\ell_{B}$, so long as Ampere's law holds $v_{\rm drift}$ cannot rapidly self-adjust, so the instabilities can only saturate non-linearly with $W\sim 1$, i.e.\ by inducing an effective damping (hence collision or scattering rate) of order the growth rate: $\omega_{\delta} \sim \omega_{\rm max}\,v_{\rm drift}/v_{\rm crit}$. This therefore must enhance the Ohmic term, which allows ${\bf B}$ to diffuse, to reduce the current and $v_{\rm drift}$ until the growth can be shut down by achieving $v_{\rm drift} \lesssim v_{\rm crit}$. }

While the details of the anomalous scalings are uncertain and complicated, as shown in \S~\ref{sec:anomalous.eta}, if we insert any of the heuristic scalings above into our three-species model in \S~\ref{sec:deriv} (e.g.\ modify the anisotropic pressure terms, account for various effective Reynolds-averaged terms from fluctuations, or allow for effective anomalous collisional damping), we obtain the same qualitative scaling. In each case Eq.~\ref{eqn:nonideal.classical} remains valid with the ambipolar and Hall terms unmodified to leading order ($\eta_{H} \rightarrow \eta_{H}^{0}$, $\eta_{A} \rightarrow \eta_{A}^{0}$), but $\eta_{O} \rightarrow \eta_{O}^{0} + \eta_{O}^{\rm an}$ (i.e.\ the anomalous term adds linearly in $\eta_{O}$), where dimensionally $\eta_{O} \sim (c^{2}\,m_{-}/4\pi\,q_{-}^{2} n_{-})\,\omega^{\rm an}\,(1 + v^{2}_{\rm drift}/v^{2}_{T})^{2}$ for $v_{\rm drift} \gtrsim v_{T}$. 

Briefly, we note here and in \S~\ref{sec:anomalous.eta} that while analogous instabilities exist when the neutral-charge carrier ``slip'' velocity (as opposed to the current drift velocity) becomes superthermal \citep[with $v_{\rm drift}$ still subthermal; see][]{tytarenko:two.fluid.drift.instabilities,squire.hopkins:RDI}, it is less clear what their non-linear outcomes should be. This is partially for lack of study, but also because they can be stabilized on microscales by pressure effects in the neutral+ionized fluids, and some particle-based simulations of the RDI have argued that saturation can involve grains ``running away'' (i.e.\ dust-gas separation rather than stronger coupling; \citealt{moseley:2018.acoustic.rdi.sims}).

\subsubsection{Other Assumptions}
\label{sec:superthermal:other}

Note that even in the superthermal limit, assumption {\bf (1)} from \S~\ref{sec:deriv.nonideal.mhd}, that displacement currents can be safely neglected, is still usually valid (producing a large displacement current would require much larger drift speeds, which we will show below is prevented by the effects discussed in \S~\ref{sec:superthermal:collisionrates}-\ref{sec:superthermal:instabilities} in the regimes of interest). We briefly note that the effect of violating {\bf (1)} would be to decouple $B$ so it detaches and propagates at $c$, akin to an even much larger anomalous resistivity. 
In {\bf (2)}, the neglect of the battery terms, which requires $a_{\rm ext} \ll \omega_{i} v_{\rm drift}$, is not invalidated by superthermal drift (so long as it was already valid with sub-thermal drift, its validity should only be strengthened by superthermal drift). 
Assumption {\bf (3)}, while common, is not strictly necessary for Eq.~\ref{eqn:nonideal.classical} -- one can simply include the appropriate coefficients $\omega^{{\rm ion/rec},0}$ and $\omega^{\rm coll}_{-+}$ in $\omega_{i}^{0}$. 

\subsection{An Approximate Treatment in the Fast Drift Limit}
\label{sec:deriv.nonideal.general}

\subsubsection{Corrected Collision Rates}

In the spirit of remaining as close as possible to the traditional non-ideal MHD formulation of \S~\ref{sec:deriv.nonideal.mhd}, note that (excepting trivial cases) $\bar{\bf J}$, $\bar{\bf J}\times \bar{\bf B}$, $\bar{\bf J}\times \bar{\bf B}\times \bar{\bf B}$ form a complete basis so it is technically possible to express the information in Eq.~\ref{eqn:induction.full}  via a set of ``effective'' coefficients:
\begin{align}
\label{eqn:nonideal.eff} \partial_{t}\bar{\bf B} &\approx \nabla \times \left[   \bar{\bf U} \times \bar{\bf B}  - \eta^{\rm eff}_{O}\,\tilde{\bf J}_{A} - \frac{\eta^{\rm eff}_{H}}{\bar{B}}\tilde{\bf J}_{A}\times \bar{\bf B}  + \frac{\eta^{\rm eff}_{A}}{\bar{B}^{2}} \tilde{\bf J}_{A} \times \bar{\bf B} \times \bar{\bf B} \right] \ , 
\end{align}
where the coefficients $\eta$ can be dimensionally expressed as: 
\begin{align}
\label{eqn:etas.general} \eta_{O}^{\rm eff} &\equiv \frac{c^{2} a_{O}^{\rm eff} }{4\pi}
= \frac{c \bar{B}}{4 \pi |q_{-}| \bar{n}_{-}} \frac{\omega_{\rm coll}^{\rm eff,\,O}}{\bar{\Omega}_{-}} \ , \\ 
\nonumber \eta_{H}^{\rm eff} &\equiv \frac{c^{2} a_{H}^{\rm eff} }{4\pi}
= \frac{c \bar{B}}{4 \pi |q_{-}| \bar{n}_{-}} \mathcal{F}^{\rm eff}_{H} \ , \\ 
\nonumber \eta_{A}^{\rm eff} &\equiv \frac{c^{2} a_{A}^{\rm eff} }{4\pi}
= \frac{c \bar{B}}{4 \pi |q_{-}| \bar{n}_{-}} \frac{\bar{\Omega}_{+}}{\omega_{\rm coll}^{\rm eff,\,A}} \ .
\end{align}
Here $\mathcal{F}_{H}^{\rm eff} \equiv c\,a_{H}^{\rm eff}  |q_{-}| \bar{n}_{-}$ is a dimensionless $\mathcal{O}(1)$ function and $\omega_{\rm coll}^{\rm eff,\,j}$ are  effective collision frequencies. 
In this form, we have simply re-parameterized our ignorance of the microphysics into the effective collision rates. 

Our discussion in \S~\ref{sec:superthermal} \&\ \ref{sec:deriv.more} argues that we can capture the leading-order effects of superthermal drift in our derivation (\S~\ref{sec:deriv}) by (1) modifying the effective collision rates $\omega_{\rm coll}^{0}$ by multiplying by the superthermal drift/slip dependent corrections (\S~\ref{sec:superthermal:collisionrates}, \ref{sec:epstein.deriv}), and (2) adding an appropriate anomalous resistivity which ``turns on'' at $v_{\rm drift} > v_{T}$ (\S~\ref{sec:superthermal:instabilities}, \ref{sec:anomalous.eta}). 
So our effective coefficients become: 
\begin{align}
\nonumber \eta_{O}^{\rm eff} &\rightarrow  \eta_{O}^{0}\,\left[ 1 +  \frac{v_{\rm slip,\,eff}^{2}+ |{\bf v}_{\rm drift}^{0}|^{2}}{v_{T}^{2}} \right]^{1/2} +  \eta_{O}^{\rm an} \, \Theta\left[\frac{v_{\rm drift}}{v_{T}} \right] \ , \\ 
\nonumber \eta_{O}^{\rm an} &\equiv \frac{c^{2} m_{-}}{4\pi |q_{-}|^{2} \bar{n}_{-}} \omega^{\rm an}\left[ 1 +  \frac{|{\bf v}_{\rm drift}^{0}|^{2}}{v_{T}^{2}} \right]^{1/2} \ , \\
\nonumber \eta_{H}^{\rm eff} &\rightarrow \eta_{H}^{0} \ ,  \\ 
\label{eqn:etas.general.specific}\eta_{A}^{\rm eff} &\rightarrow \eta_{A}^{0} \,\left[ 1 +  \frac{v_{\rm slip,\,eff}^{2}}{v_{T}^{2}} \right]^{-1/2} \ ,
\end{align}
where $v_{\rm slip,\,eff}^{2}/v_{T}^{2} \equiv -1/2 + \sqrt{1/4 + |{\bf v}_{\rm slip,\,\bot}^{0}|^{2}/v_{T}^{2}}$ and $v_{\rm drift,\,max}^{2} \equiv v_{\rm slip,\,eff}^{2}+ |{\bf v}_{\rm drift}^{0}|^{2}$ (with ${\bf v}_{\rm slip,\,\bot}^{0}$ and ${\bf v}_{\rm drift}^{0}$ defined in Eq.~\ref{eqn:vdrift}, using $\eta_{A}^{0}$ -- i.e.\ from the usual coefficients without velocity-dependence). 
Here $\Theta(x)$ can be the Heaviside step function or any similar function (e.g.\ $e^{-1/x}$). 
It is not necessary to drop inelastic reactions in the $\eta^{0}$ terms if they are important. Motivated by the extensive studies discussed \S~\ref{sec:superthermal}, we take $\omega^{\rm an}$ to be the faster of either the gyro or plasma frequencies, e.g.\ $\omega^{\rm an} \sim (\bar{\Omega}_{-}^{2} + \omega_{p-}^{2})^{1/2}$, but we will show that the exact value does not matter so long as it reflects a sufficiently fast frequency of the system. Eq.~\ref{eqn:etas.general.specific} is approximate: in \S~\ref{sec:extra.multi} we discuss how more accurate correction terms can be obtained numerically, but show Eq.~\ref{eqn:etas.general.specific} captures their important asymptotic scalings.

Prescriptions akin to this are already used in laboratory, Solar, and ionospheric plasma physics applications (see references in \S~\ref{sec:superthermal:instabilities} and \citealt{somov:2000.anomalous.resistivity.prescriptions.solar.sims,roussev:2002.anomalous.resistivity.prescriptions.summary,ni:2007.cowling.conductivity.vs.sweet.parker.and.anomalous,farder:2023.anomalous.resistivity.models.solar.review}, for examples). 
However, Eq.~\ref{eqn:etas.general.specific} expands upon these (based on our discussion above) by 
(1) including Ohmic, Hall, and ambipolar terms (as opposed to just Ohmic); 
(2) including the drift-dependent corrections to the $\eta_{0}$ terms (usually ignored); 
(3) generalizing $\eta^{\rm an}$ to allow both weakly ($\Omega_{e} < \omega_{pe}$, assumed in those papers) and strongly ($\Omega_{e} > \omega_{pe}$) magnetized limits; 
(4) allowing for mostly-neutral gas (as compared to just fully-ionized fluids); 
(5) generalizing (\S~\ref{sec:multispecies} \&\ \ref{sec:extra.multi} below) to multi-species cases; 
and (6) providing criteria for the critical drift/slip speeds in different media.

Likewise, Eq.~\ref{eqn:etas.general.specific}  is analogous to widely adopted treatments of similar problems in astrophysical applications of thermal conduction and viscosity, where naively applying the ``classical'' conduction/viscosity equations \citep{spitzer:conductivity,braginskii:viscosity} can produce sharp gradients that imply unphysically superthermal drift velocities (of e.g.\ thermal electrons or ions in the ``saturated'' limit; e.g.\ \citealt{spitzer:1962.ionized.gases.book,chapman:1970.gas.dynamics.book,cowie:1977.evaporation}). It has also been widely recognized that for high-$\beta^{\rm plasma}$ plasmas, the maximum anisotropy of the velocity distribution function (which can be considered similarly to a maximum drift speed) is limited by microscopic instabilities like the whistler, mirror, and firehose, which are activated above certain thresholds (
\citealt{Schekochihin2008b,komarov:heat.flux.suppression.ICM,komarov:conduction.vs.mirror.instab,komarov:whistler.instability.limiting.transport,kunz:firehose,squire:2016.alfvenic.perturbations.limits.anisotropy,squire:2017.max.braginskii.scalings,squire:2017.max.anisotropy.kinetic.mhd}). These are similarly approximated on macroscopic scales by multiplying the ``classical'' coefficients by corrections designed to represent some anomalous scattering that limits the velocity anisotropy  \citep[see e.g.][]{riquelme:viscosity.limits,roberg:2016.whistler.turb.conduction.suppression,roberg:2018.whistler.turb.conduction.suppression,su:2016.weak.mhd.cond.visc.turbdiff.fx,squire:2017.kinetic.mhd.alfven,hopkins:cr.mhd.fire2}. We are simply pointing out that similar terms should apply in the other astrophysical cases of interest here (as was also argued by authors like \citealt{norman.heyvaerts:1985.anomolous.resistivity.current.instabilities.review}), generalizing them to cases with e.g.\ appreciable neutrals and non-electron-ion plasmas, while also accounting for similar the relevant terms in the ambipolar drift.

\subsubsection{Generalization To Arbitrary Numbers of Species}
\label{sec:multispecies}

\begin{figure}
	\centering\includegraphics[width=0.95\columnwidth]{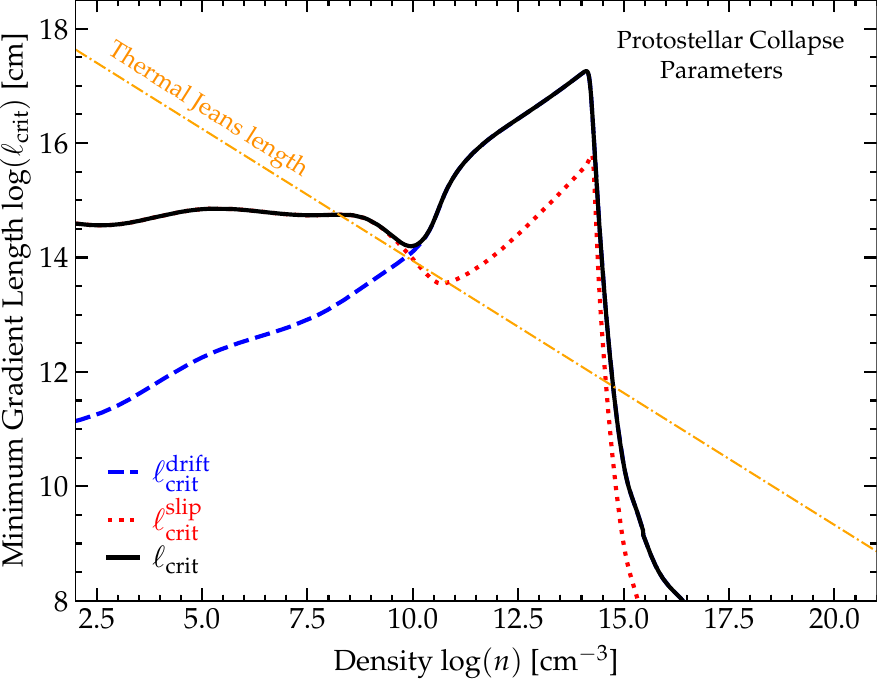} 
	\centering\includegraphics[width=0.95\columnwidth]{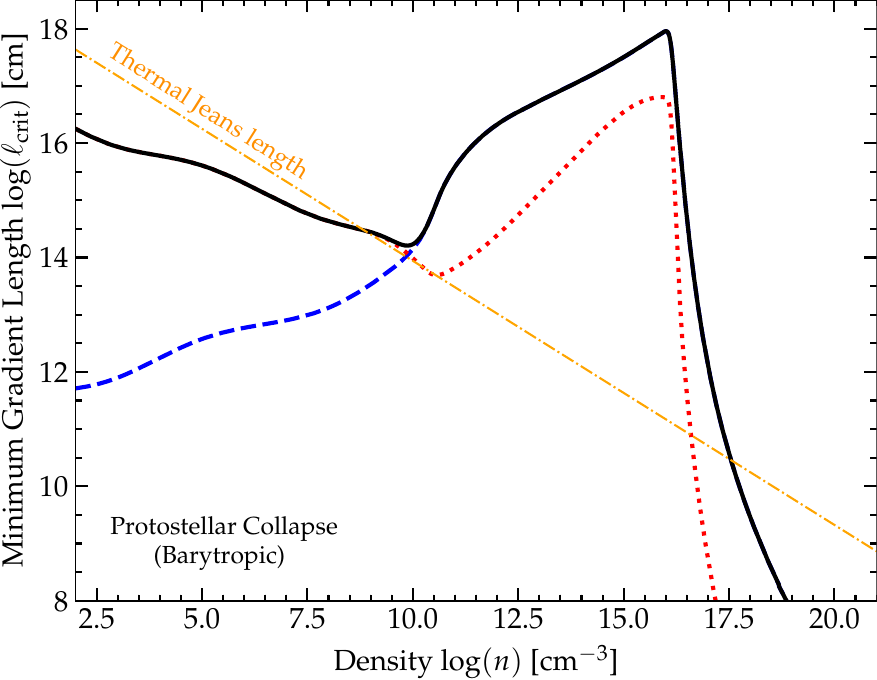} 
	\centering\includegraphics[width=0.95\columnwidth]{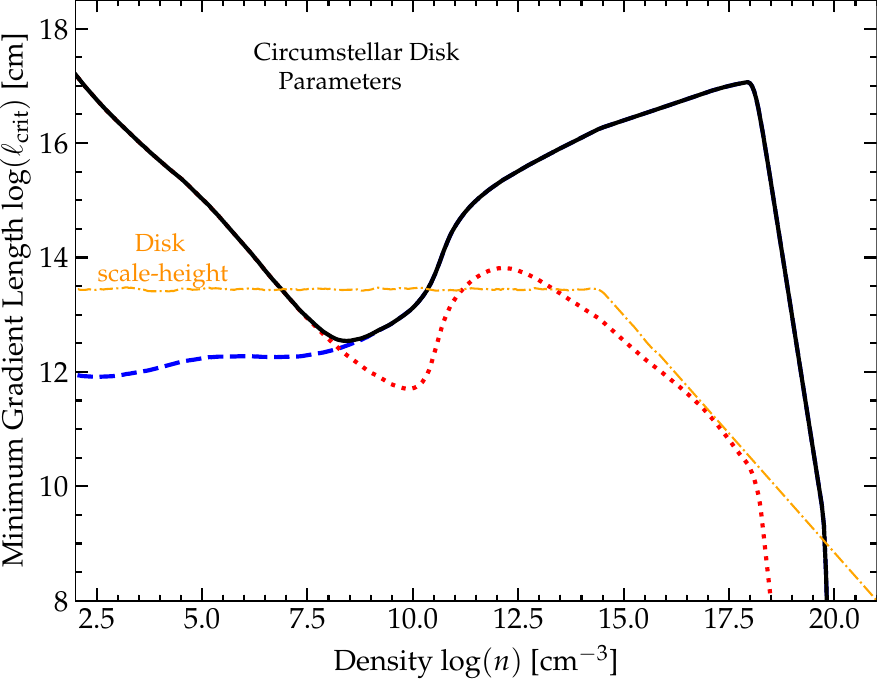} 
	\caption{Critical magnetic gradient scale-length $\ell_{\rm crit}$, below which the drift ($\ell_{\rm crit}^{\rm drift}$) or slip ($\ell_{\rm crit}^{\rm slip}$) speeds become superthermal, versus density $n$ for the same three model variants as Fig.~\ref{fig:etas}. Strong gradients below this scale will be rapidly erased by enhanced resistivity (Fig.~\ref{fig:etas}; \S~\ref{sec:consequences}). For comparison we plot the disk scale-height or thermal-pressure scale-length $h_{T}$ at each $n$ in the models. At circumstellar disk radii from $\sim 0.1-100\,$au (densities $n \sim 10^{10}-10^{18} {\rm cm^{-3}}$), and at densities where the Hall diffusion would naively dominate absent anomalous correction terms (for $v_{\rm drift} \ll v_{T}$), we have $\ell_{\rm crit} \gg h_{T}$, so magnetic structures and Hall effects on disk scales will be strongly suppressed. 
	\label{fig:lcrit}}
\end{figure}

Given the full set of restricted assumptions in \S~\ref{sec:deriv.nonideal.mhd}, one can derive coefficients $\eta^{0}_{O,H,A}$ for an arbitrary number of species. In principle, as outlined in more detail in Appendix~\ref{sec:extra.multi}, one could numerically solve an entire chemical network for species abundances and simultaneously obtain the effective coefficients $\eta_{O,H,A}$ for an arbitrary set of velocity-dependent collision rates and anomalous resistivities $\omega^{\rm an}_{ij}$, based on the relative velocities between all pairs of species. But this would involve fundamental changes to the sorts of chemical networks and solvers usually employed, as well as introducing a number of ambiguities. Instead, much of the attention in modeling weakly ionized systems has focused on using chemical networks  to calculate different charged-particle abundances $n_{i}$, which are then used in a fully operator-split manner to calculate $\eta^{0}_{O,H,A}$ from simplified expressions that depend only on $n_{i}$ (and background properties like $T$ or $B$, but not drift speeds). So it is important to consider how to generalize our approach to include many species. In this section, we develop a prescription that captures anomalous resistivity effects with sufficiently simple closed-form expressions that can be straightforwardly applied to the outputs of standard chemical-network methods (i.e.\ using information already supplied by those codes, rather than modifying the chemical computation itself, as we discuss in Appendix~\ref{sec:extra.multi}, where we validate these expressions against the numerical approach). 

Per Appendix~\ref{sec:extra.multi}, for the applications of interest here (weakly ionized systems), the ``standard'' coefficients are derived in the multi-species limit by taking Eq.~\ref{eqn:mom}, neglecting all terms except Lorentz and neutral collisions, and assuming these are in equilibrium, to solve for $\delta{\bf u}_{j} \equiv {\bf u}_{j}-{\bf u}_{n}$ as a function of ${\bf E}^{\prime}\equiv {\bf E} - {\bf u}_{n} \times {\bf B}/c$ with $q_{j}({\bf E}^{\prime} + \delta{\bf u}_{j} \times {\bf B}/c) = m_{j}\omega^{\rm coll}_{jn} \delta{\bf u}_{j}$ (this is the local ``terminal velocity approximation''). This leads to the well-known result that the relative speeds of different particles scale with their Hall parameter $\propto q_{j}/m_{j}\omega^{\rm coll}_{j}$, often parameterized as:
\begin{align}
\label{eqn:beta} \tilde{\beta}_{j} &\equiv \frac{q_{j} B}{m_{j} c \gamma_{j} \rho} 
\end{align}
\citep{cowling:1976.mhd.book,nakano:1986.nonideal.mhd.formulation.review,wardle.ng:1999.nonideal.coefficients}. The contribution of each to the current is then $|\delta {\bf J}| \propto n_{j} q_{j} \tilde{\beta}_{j}$.\footnote{Note since this only appears as a weight function, the choice to multiply by $e B/c\rho$ is arbitrary. One can generalize Eq.~\ref{eqn:beta} further for the case with arbitrary neutral density, where for the fully-ionized case the weights $\tilde{\beta}_{j} \propto q_{j}/m_{j}$, but in this case the various classical $\eta$ terms are generally negligible compared to the ideal MHD terms.}
Consider e.g.\ the simple case where there are two negative charge carriers -- one ``light'' and relatively weakly collisionally coupled, and one ``heavy'' and much more strongly collisionally coupled (both with $q \sim -e$).  This is often the case with electrons and charged grains, where the assumption of such a mass hierarchy is well justified. This gives the intuitive result that even if $n^{-}_{\rm grains} > n_{e}$, the current is still primarily carried by the electrons, unless $n^{-}_{\rm grains}/n_{e} \gg (m_{\rm grain} \gamma_{\rm grain})/(m_{e} \gamma_{en}) \sim 10^{9.3} (R_{\rm grain}/{\rm \mu m})^{2}$. For the systems of interest (e.g.\ protostellar cores, circumstellar/protostellar disks, planetary atmospheres), the latter condition is rarely met, so even though grains can dominate the total charge, they do not actually carry most of the current. This means the current is being carried by relatively fewer free electrons, implying even higher drift speeds.\footnote{Consider, for example, the limit of extremely large grains or ``rocks.'' Even if these somehow accumulated free electrons, clearly their electromotive forces are negligible and they would remain anchored to their positions, with Coulomb forces accumulating an opposite charge ``locked'' to their location. Thus even if they do not recombine, it is more appropriate to consider these effectively part of the neutral background, while the current and MHD dynamics of the fluid on scales large compared to the inter-grain separation are carried entirely by the remaining free electrons.}

In that limit, the ``standard'' Ohmic coefficient then scales as $\eta_{O}^{0} = (c B/4\pi)/(\sum n_{j} q_{j} \tilde{\beta}_{j})$, i.e.\ is determined by the current-carrying species (the species which contributes the largest $|\delta {\bf J}|$). To determine an effective ${\bf v}_{\rm drift}$ of the dominant current carriers then, in order to estimate an effective Ohmic correction, we do not need to consider the relative drift velocity of every possible pair of charged species (thousands or millions of possible combinations), but the same weighted average. We therefore take Eq.~\ref{eqn:vdrift} (${\bf v}_{\rm drift} = (c \nabla \times \bar{\bf B})/(4\pi \bar{n}_{+}q_{+}) = (c \nabla \times \bar{\bf B})/(4\pi \bar{n}_{-} |q_{-}|)$) and replace it with the weighted version: 
\begin{align}
\label{eqn:nq} {\bf v}_{\rm drift} &\rightarrow \frac{c \nabla \times \bar{\bf B}}{4\pi \langle \bar{n}_{-}|q_{-}| \rangle},   \\ 
\nonumber \langle \bar{n}_{-}|q_{-}| \rangle &= \langle\bar{n}_{+} q_{+}\rangle  \rightarrow \frac{\sum_{j} n_{j} |q_{j}| |\tilde{\beta}_{j}|} {\sum_{j} |\tilde{\beta}_{j}|} \ .
\end{align}
Meanwhile $v_{\rm slip,\,\bot}$ (and hence $v_{\rm drift,\,max}$) can be calculated directly from Eq.~\ref{eqn:vdrift} as ${\bf v}_{\rm slip,\,\bot} = (\bar{\rho}/\bar{\rho}_{n})\,(\eta_{A}^{0}/\bar{B}^{2}) \nabla\times\bar{\bf B}\times\bar{\bf B}$ (unchanged), using the total density and total density of (all) neutrals. Since this builds in the weighting for multiple species already present in $\eta_{O}$ (defined in \S~\ref{sec:extra.multi}), this again ensures the proper weight (here towards slip velocity between the neutrals and the important inertia-bearing charged species). 
Note that Eq.~\ref{eqn:nq} immediately reduces to our expression from \S~\ref{sec:deriv.nonideal.mhd} $\langle \bar{n}_{-}|q_{-}| \rangle = n_{-}|q_{-}| = n_{+}q_{+}$ when we just have two dominant charge carriers (one positive one negative), as it should. We similarly replace $n_{-} |q_{-}|$ or $n_{i} q_{i}$ by $\langle \bar{n}_{-} |q_{-}| \rangle$ (Eq.~\ref{eqn:nq}) in definitions like Eq.~\ref{eqn:lcrit} for $\ell_{\rm crit}$. 

For $\eta_{O}^{\rm an}$, the salient anomalous resistivity is that of the same species that dominates the weighted average in our drift/current/Ohmic expression, so we can write: 
\begin{align}
\label{eqn:eta.an} \eta_{O}^{\rm an} \sim \frac{c \bar{B}}{4\pi \langle \bar{n}_{-} |q_{-}|\rangle} \left[1 + \frac{4\pi \langle \bar{\rho}_{-} \rangle c^{2}}{\bar{B}^{2}} \right]^{1/2} \ .
\end{align}
This simply makes the same replacement for $\bar{n}_{-} q_{-} \rightarrow \langle \bar{n}_{-} |q_{-}|\rangle \sim e n_{e}$ in the prefactor, and defines $\bar{\rho}_{-} \equiv m_{-} \langle n_{-} \rangle \sim m_{e} n_{e}$ to reflect the same dominant current-carrying species as $\langle \bar{n}_{-} |q_{-}|\rangle$. 
Also note that in the multi-species treatment, the pre-factors $\propto (n_{-}|q_{-}|)^{-1}$ in Eq.~\ref{eqn:etas.general} are replaced by a similar weighted factor, so the two are consistent.

The generalization of $v_{T}$ as a ``threshold'' for anomalous behavior for the multi-species case is slightly less obvious. Of course, per \S~\ref{sec:superthermal} the exact correct value here depends on the exact parameters and microphysics by way of exactly which instabilities and collision and interaction terms dominate on which scales. We can generically define an effective isothermal soundspeed for a multi-component plasma as: 
\begin{align}
\label{eqn:vT} v_{T}^{2} &\equiv \frac{P_{\rm eff} }{\rho_{\rm eff}} \equiv \frac{\sum_{j} \psi_{j} n_{j} k_{B} T_{j}}{\sum_{j} \psi_{j} m_{j} n_{j}} \ ,
\end{align}
where $\psi_{j}$ is some weight. In most applications of interest here, since it is already assumed in the standard formulation, we can take $T_{j} = T$ (a single temperature system). 

Naively, a sensible weight might be $\psi_{j} \sim $\,constant for charged species and $=0$ for neutrals, or $\psi_{j} = |Z_{j}| = |q_{j}/e|$. For an ion-electron plasma, this would give the expected answer with $v_{T}$ the thermal-ion speed (up to an $\mathcal{O}(1)$ prefactor, which is degenerate with our uncertainties in the ``most salient'' $v_{T}$ in any case). But this choice of $\psi$ would not correctly capture the case where one or both carriers are dust grains: with that definition $v_{T} \sim (k_{B} T / m_{\rm grain})^{1/2} \sim 0.01\,{\rm cm\,s^{-1}}\,(T/10\,{\rm K})^{1/2}\,(a_{\rm grain}/{\rm \mu m})^{-3/2}$ would become vanishingly small. This produces a number of results that might technically be valid if the {\em only} species were dust grains (i.e.\ no neutrals or ions or electrons present -- an extreme ``dusty plasma'' case) but do not make sense in the context of interest here (where the dust thermal speed is extremely slow compared to that of other important species in the plasma). We can formulate a $\psi$ to capture realistic dusty plasmas by noting that in the limits of interest where dust is a dominant charge carrier, we are always in the weakly-ionized limit (large neutral density) with the dust strongly coupled to said neutrals ($\omega^{\rm coll}_{{\rm dust},n} \gg \Omega_{{\rm dust}}$). This means the neutrals should be included in Eq.~\ref{eqn:vT} since their inertia modifies the effective wavespeed of perturbations to the dust, so $v_{T} \sim v_{T,n}$, and instabilities are excited when $v_{\rm drift}$ exceeds the neutral thermal speed. This also accords with the formal analysis of dust streaming instabilities/RDIs in \citet{squire.hopkins:RDI,squire:rdi.ppd,hopkins:2017.acoustic.RDI}, where $v_{T,n}$ is the speed grains must be moving to excite strong instabilties on small scales. We can capture these cases approximately by taking:
\begin{align}
\psi_{j} &\equiv \begin{cases}
|q_{j}/e| & (q_{j} \ne 0) \\ 
\frac{1}{\sum_{i} |q_{i}| n_{i}}\sum_{i} \frac{\omega^{\rm coll}_{ij} |q_{i}| n_{i}}{\omega^{\rm coll}_{ij} + \Omega_{i}}  & (q_{j} = 0) \ .
\end{cases}
\end{align}
Note the latter sum is nearly identical to sums that already need to be computed for $\eta_{O,H,A}^{0}$ in the standard multi-species formalism \citep{wardle.ng:1999.nonideal.coefficients} so entails no additional computational expense. 
This captures the intuitive behavior in the limits of interest (and, as we show in \S~\ref{sec:extra.multi}, the numerical behavior in more exact computations): in the weakly-ionized limit with any modest neutral coupling, this will tend to $v_{T} \rightarrow v_{T,n}$, while in the strongly ionized limit, this will tend to the thermal ion speed, as desired.

As noted above, another critical wavespeed of interest is the effective \Alf\ speed $v_{A}^{\rm eff}$ (with various instabilities excited for $v_{\rm drift} > v_{A}^{\rm eff}; \S~\ref{sec:superthermal:instabilities})$, so we should more generally take
\begin{align}
\nonumber v_{T} &\rightarrow {\rm MIN}\left[ v_{T}^{\rm therm} \, , \, v_{A}^{\rm eff} \right] \\ 
\label{eqn:vA.eff} v_{A}^{\rm eff} &\equiv \frac{B}{\sqrt{4\pi \sum_{j} \psi_{j} m_{j} n_{j} }}
\end{align}
where $v_{T}^{\rm therm}$ is given by Eq.~\ref{eqn:vT} and the weights in $v_{A}^{\rm eff}$ follow from the same argument as above for $v_{T}^{\rm therm}$. Qualitatively, most super-\Alf{ic} instabilities are similar to non-relativistic CR streaming instabilities \citep{wentzel:1968.mhd.wave.cr.coupling} with growth rates $\sim \Omega_{e} (v_{\rm drift}/v_{A} - 1)$. If all species have $\omega^{\rm coll} \gg \Omega$, these instabilities and the excited \Alf\ waves will be strongly-damped, but this case is well into the classical Ohmic regime where the anomalous resistivity would be relatively small anyway. If all species have $\omega^{\rm coll} \ll \Omega$, i.e.\ the classical ambipolar regime, then $v_{A} \rightarrow B/\sqrt{4\pi\rho_{i}}$ becomes the ``ion \Alf\ speed,'' but this is $\gg v_{T}$ in weakly-ionized systems. But in the classical Hall regime ($\Omega_{+} \ll \omega \ll \Omega_{-}$), this gives interesting behavior: the effective \Alf\ speed is the ideal $v_{A}^{\rm eff} \approx B/\sqrt{4\pi\rho}$ because the wavelengths of interest are sufficiently large that the \Alf\ wave frequency is much smaller than ion-neutral collisions, but the growth rate for the fast carriers (e.g.\ electrons) is much larger than the damping/collision rates. In this limit, $v_{A}^{\rm eff} < v_{T}^{\rm therm}$ is equivalent to $\beta^{\rm plasma} > 1$. So for systems in the classical Hall regime, with $\beta^{\rm plasma} > 1$, the inclusion of $v_{A}^{\rm eff}$ in $v_{T}$ can be a non-negligible correction. While, for the examples we plot below this makes almost no difference to the results, the most interesting consequence in this limit is that $\ell_{\rm crit}^{\rm drift} \rightarrow 73 {\rm au} (n / 10^{15}{\rm cm^{-3}})^{1/2} (n_{e}/{\rm cm^{-3}})^{-1}$, independent of $B$ or $T$ and dependent only on the total density and free electron density of the system.

\subsubsection{Specific Examples for Calculation}
\label{sec:plots}

In Figs.~\ref{fig:regimes}, \ref{fig:etas}, \&\ \ref{fig:lcrit}, we illustrate the scalings derived above, showing relative values of $\eta_{O,\,H,\,A}$ and $\ell_{\rm crit}$ in some simple examples. In order to calculate specific values, we require a specific model for the chemistry (relative abundances of different charged species), density, temperature, magnetic field strength, etc. 
To do so, we consider a few different ``default'' models taken from the NICIL code \citep{wurster:2021.ambipolar.small.fx.starform.hall.fx.larger}, a widely-used chemical network designed to compute the coefficients $\eta_{O,\,H,\,A}^{0}$ in mostly-neutral gas, with a large number of different ionized species, as well as both positively and negatively charged grains obeying an assumed grain size distribution. The coefficients in NICIL are calculated using the standard assumptions in \S~\ref{sec:deriv.nonideal.mhd}, notably assuming negligible drift, so we compare them to the corrected coefficients and scalings with our expressions, using the values of $T$, $B$, $n_{j}$, etc.\ output by the code.
Specifically we consider (1) their ``non-ideal MHD protostellar collapse'' model and (2) their ``baryotropic equation-of-state collapse'' model \citep{wurster:2019.non.ideal.rad.mhd.disk.sims}, and (3) a ``protoplanetary disk'' model (with a \citealt{mathis:1977.grain.sizes}-like grain size distribution, attenuated cosmic ray ionization rate, and disk extended from $\sim 0.01-300$\,au; \citealt{wurster:2021.ambipolar.small.fx.starform.hall.fx.larger}). These make different assumptions for the scalings of $T$, $B$, and ionization fractions with density so give different coefficients relevant in different regimes. We stress that we are not arguing for any particular chemical model or scenario among these, but simply using them to illustrate the qualitative, systematic effects of superthermal drift in different situations. To this end, we have also re-computed Figs.~\ref{fig:etas}-\ref{fig:lcrit} using a much more simplified five-species chemical model described in Appendix~\ref{sec:extra.multi}. That gives different absolute values of the coefficients $\eta_{O,\,H,\,A}$, but the qualitative conclusions and systematic effects of the corrections from superthermal drift are the same in all cases.

As noted, NICIL considers a large number of species, with e.g.\ charged dust grains and many different ions relevant in dense gas. Thus our simple three-species model alone in \S~\ref{sec:deriv} does not un-ambiguously define ``which'' values of $v_{T}$ and other salient quantities to use in calculating the correction terms. For these, we specifically follow the generalized multi-species equations in \S~\ref{sec:multispecies} above (Eqs.~\ref{eqn:eta.an}-\ref{eqn:vT} for the effective $\eta_{O}^{\rm an}$ and $v_{T}$, respectively), with the dust included alongside the gas using the same grain sizes assumed in the NICIL calculation.

In Fig.~\ref{fig:regimes}, we estimate which of the Ohmic/Hall/ambipolar coefficients is largest in magnitude as a function of $B$ and $n$ (taking just a simple model for $T(n)$ from \citet{machida:2006.second.core.form.simplified.eos.sims} and allowing $B$ to freely vary at fixed values of all other properties at each $n$, as our inputs to NICIL). In Fig.~\ref{fig:etas} and Fig.~\ref{fig:lcrit}, we show the values of $\eta_{O,H,A}$ and the critical gradient length scales $\ell_{\rm crit}^{\rm drift}$ and $\ell_{\rm crit}^{\rm slip}$ (Eq.~\ref{eqn:lcrit}) for each of the three models above. In Figs.~\ref{fig:regimes} \&\ \ref{fig:etas}, we compare the standard non-ideal formulation for $\eta_{O,H,A}=\eta_{O,H,A}^{0}$ (Eq.~\ref{eqn:nonideal.classical}), which ignores anomalous corrections and is therefore only valid when $\ell_{B} \gg \ell_{\rm crit}$ ($v_{\rm drift} \ll v_{T}$), to the corrected coefficients for the case of an $\ell_{B}$ which would otherwise give $v_{\rm drift} \sim 10\,v_{T}$ (using the implied $v_{\rm drift}$ from the standard formulation, Eqs.~\ref{eqn:vdrift} \&\ \ref{eqn:nq}), i.e.\ $\ell_{B} \sim 0.1\ell_{\rm crit}^{\rm drift}$. So Fig.~\ref{fig:lcrit} plots the dividing line between these cases, in terms of $\ell_{B}$.

\section{Where Might the ``Standard'' Formulation Be Problematic?}
\label{sec:where}

Where might the correction terms in \S~\ref{sec:deriv.nonideal.general} be important? Consider the case with $v_{\rm drift} > v_{\rm slip}$, and note we can write $\ell^{\rm drift}_{\rm crit} \approx (v_{A,i}/v_{T}) d_{i}$, 
where $v_{A,i}\equiv B/\sqrt{4\pi\rho_{i}}$ and $d_{i}\equiv c/\omega_{pi} = v_{A,i}/\Omega_{c,i} = r_{g,i}/\sqrt{\beta^{\rm plasma}_{i}}$ are the ion \Alf\ speed and inertial length (and $\beta^{\rm plasma}_{i} \equiv (c_{s}/v_{A,\,i})^{2} \equiv P_{{\rm therm},\,i}/2\,P_{B}  \equiv 4\pi\,n_{i}\,k_{B}\,T_{i}/B^{2} \equiv f_{i}\,\beta^{\rm plasma}$ is the ion plasma $\beta^{\rm plasma}_{i}$, with $r_{g,i}$ the ion gyro radius). 
This means that for well-ionized plasmas, so long as $\ell_{B} \gg d_{i}$ (for $v_{T}\sim v_{A}$) or $\gg d_{i}/\beta_{\rm plasma}^{1/2}=r_{g}/\beta_{\rm plasma}$ (for $v_{T} \sim v_{\rm thermal}$), the system will be ``safely'' in the $v_{\rm drift} \ll v_{T,\,i}$ regime.\footnote{For example, in the diffuse interstellar, circumgalactic, and intergalactic medium, stellar photospheres, HII regions, and OB winds, the ratio of typical $\ell_{B}$ for order-unity changes in $B$ (i.e.\ the \Alf\ scale of turbulence) to $\ell_{\rm crit}$ is $\sim 10^{10}-10^{15}$, and it reaches $\gg 10^{20}$ in stellar interiors. Even in extremely low-$\beta^{\rm plasma}$ environments (as low as $< 10^{-6}$), such as those proposed for quasar accretion disks in \citet{hopkins:superzoom.overview,hopkins:superzoom.analytic,hopkins:superzoom.disk} this ratio is still $\gg 10^{9}$.} 

However, in weakly-ionized plasmas such as dense molecular cloud cores, circum-stellar and circum-planetary disks (including protostellar and protoplanetary disks), planetary atmospheres, and some cool-star outflows, $f_{i}$ can become extremely small (e.g.\ $\sim 10^{-15}$; \citealt{galli:1993.ambipolar.diff.for.star.formation,basu:1994.nonideal.mhd.formulation.collapse.models,kunz:2010.sf.model.nonideal.results,dapp:2010.nonideal.mhd.braking.catastrophe}). For example, scaling to parameters expected at distances of order $\sim 1\,$au in a circumstellar disk, we have $\ell_{\rm crit} \sim 400\,{\rm au}\,(B/{\rm G})\,(0.1\,{\rm km\,s^{-1}}/v_{T,\,i})\,(0.1\,{\rm cm^{-3}}/n_{i})$ (or even larger, in the specific regime noted in \S~\ref{sec:multispecies} where $v_{A}<v_{T}^{\rm therm}$), much larger than disk scale-heights $\sim 0.05\,$au. The values of $\ell_{\rm crit}$ one might expect at different densities are illustrated more directly in Fig.~\ref{fig:lcrit}.

If one only models ``ideal'' or Braginskii/Spitzer MHD, there is nothing to prevent such strong currents from forming. However in weakly-ionized environments, the ``classical'' non-ideal terms in Eq.~\ref{eqn:nonideal.classical} (Ohmic $\eta_{O}^{0}$, Hall $\eta_{H}^{0}$, and ambipolar $\eta_{A}^{0}$) need to be accounted for. Dimensionally, the coefficients $\eta$ scale as $\mathcal{O}(\eta^{0}) \sim B\,c\,F(\boldsymbol{\Psi})/4\pi e n_{i}$ (where $F(\boldsymbol{\Psi}) \gtrsim 1$ is a dimensionless function of the different frequencies, see above), so the ratio of the non-ideal to ideal (${\bf U} \times {\bf B}$) term is $\sim F(\boldsymbol{\Psi}) \,\ell_{\rm crit}\,v_{T,\,i} / \ell_{B}\,|{\bf U}|$. Since the Ohmic and ambipolar terms are fundamentally diffusive, this means they will generally tend to mitigate against super-thermal drifts: so long as the bulk motions of the fluid are subsonic ($|{\bf U}| \ll v_{T,\,i}$), then if the drift becomes large ($\ell_{B} \lesssim \ell_{\rm crit}$), the diffusive term will become large in the induction equation, diffusing the gradient away and restoring subsonic drift speeds. However, even in the Ohmic/ambipolar regime, superthermal drifts can still arise in principle if we only include the ``classical'' terms $\eta^{0}$ and external forces (e.g.\ gravity or radiation) drive superthermal compressions/outflows/shocks. 

The problem in the classical regime is much more severe if the Hall term dominates, because its effect is fundamentally {\em not} diffusive. Indeed, simulations of Eq.~\ref{eqn:nonideal.classical} in weakly-ionized systems show that the Hall term drives the formation of thin current sheets (thinner than thermal scale lengths) causing $\ell_{B}$ to shrink to very small values at these interfaces \citep{bejarano:2011.hall.shear.instabilities,kunz:2013.hall.self.organization,lesur:2014.ppd.nonideal.effects.idealized.with.hall,bai:2017.hall.magnetic.transport.ppds}. This would thus {\em increase} $v_{\rm drift}$. And \citet{zhao:2018.hall.nonideal.mhd.corecollapse.strong.currents,zhao:2021.hall.mhd.star.formation.disk.formation} explicitly show in dense core collapse and circumstellar disk simulations (using standard non-ideal MHD formulation and coefficients $\eta^{0}_{O,H,A}$) that the implied drift velocities reach several ${\rm km\,s^{-1}}$ at temperatures $T \sim 10-100\,$K (thermal speeds $\lesssim 0.1\,{\rm km\,s^{-1}}$). So superthermal drift is clearly occurring in at least some calculations. This is illustrated in Figs.~\ref{fig:regimes}-\ref{fig:lcrit}, and discussed in \S~\ref{sec:consequences:hall} below, where we find no regime in which Hall can dominate on global scales.

A third potentially problematic regime is when dust grains become one of the primary charge carriers by absorbing free electrons (and/or ions). This can technically make the system a ``dusty plasma'' (as distinct from the more typical ``dust-laden'' plasma in most astrophysical environments; see \citealt{melzer:2021.dusty.plasmas.book,melzer:2021.magnetized.dusty.plasmas.review.book,beckers:2023.dusty.plasmas.physics.perspectives.review}). It is not always obvious that the other assumptions used in \S~\ref{sec:theory} to derive Eq.~\ref{eqn:nonideal.classical} hold in this regime, since grains can develop internal currents and fields (being complicated electrostatic and dielectric media themselves) and can undergo near continuous charge exchange reactions (so we might need to include terms like $\partial_{t}{q}_{\pm}$ which are ignored in MHD derivations), but such questions are beyond the scope of our study. Even if we simply treat grains as ``heavy'' ions (the usual approximation), their masses are enormous ($\gtrsim 10^{13}\,m_{p}$), and the ``light'' carriers are either heavy molecules or other grains. This means the effective thermal ``ion'' speeds can become extremely small and (correspondingly) they can develop extreme anisotropy in their distribution functions which usually enhances the instabilities reviewed above, allowing them to persist down to extremely small scales \citep{squire.hopkins:RDI}. Moreover, since grains are so heavy, an even greater current-carrying requirement (and therefore even higher drift speeds) is implicitly placed on the (fewer) remaining free electrons.

\section{Some Consequences}
\label{sec:consequences}

Now consider some of the important consequences of Eq.~\ref{eqn:nonideal.eff}, compared to the ``standard'' zero-drift/slip coefficients $\eta_{O,H,A}^{0}$ (Eq.~\ref{eqn:nonideal.classical}). We illustrate these in Figs.~\ref{fig:regimes}, \ref{fig:etas}, \&\ \ref{fig:lcrit}.

\subsection{Ordering and Scaling of Terms}
\label{sec:consequences:order}

Fom Eq.~\ref{eqn:etas.general} the relative importance of the non-ideal $\eta$ terms are given by ratios $\Omega_{c}/\omega_{\rm coll}$ of gyro to collision frequencies, so it is useful to consider three limits in turn, illustrated in Fig.~\ref{fig:regimes}.

The regime where ambipolar diffusion would dominate, in the ``standard'' or zero-drift-assumption limit ($\Omega_{c,\pm} \gg \omega^{0}_{\rm coll}$), generically has $v_{\rm drift} \ll v_{\rm slip}$, meaning current-driven instabilities will not be important. While there can be pure slip-driven instabilities, their effect is less clear (see \S~\ref{sec:superthermal:instabilities}). So when the slip becomes superthermal, we just have the Epstein-type correction for $\langle \sigma v \rangle$ accounting for the superthermal relative velocity of charge carriers versus neutrals in the collision rates, which decreases $\eta_{A}$ and increases $\eta_{O}$, but only by a linear factor of $v_{\rm drift/slip}/v_{T}$. Thus it will not usually change the relative ordering of the non-ideal terms (unless they are already similar in magnitude).

In the regime where Ohmic resistivity would strongly dominate with zero drift ($\Omega_{c,\pm} \ll \omega_{\rm coll}$), $v_{\rm drift} \gg v_{\rm slip}$. If the drift becomes superthermal, current-driven instabilities could appear but in this limit, the anomalous resistivity $\eta_{O}^{\rm an}  \sim \eta_{O}^{0} (\Omega_{c-,p-}/\omega_{\rm coll}^{0}) \ll \eta_{O}^{0}$ would be relatively small because the collision frequency driving the resistivity is already the fastest frequency of interest. So even if collisions strongly damp these instabilities in this limit, it makes no practical difference. The dominant correction will again simply be the proper accounting for superthermal drift in the Epstein correction to the collision rates, boosting $\eta_{O}$ by a factor of $\sim v_{\rm drift}/v_{T}$.

The regime where the Hall term would dominate with zero drift ($\Omega_{c,+} \ll \omega_{\rm coll} \ll \Omega_{c,-}$, $v_{\rm drift} \sim v_{\rm slip,-} \gg v_{\rm slip,+}$) is perhaps most interesting, despite the fact that the Hall term {\em itself} is independent of these correction factors to leading order (being independent of the effective collision frequency except for an $\mathcal{O}(1)$ prefactor that depends only on the dimensionless ratios of different collision rates). In this regime, by definition, the zero-drift Ohmic term is relatively small $\eta_{O}^{0} \sim |\eta_{H}^{0}|\,\omega_{\rm coll}/\Omega_{c-} \ll |\eta_{H}^{0}|$. But if the drift becomes superthermal, current-driven instabilities will appear and be weakly or negligibly damped. Thus $\eta_{O} \rightarrow (\eta_{O}^{0}+\eta_{O}^{\rm an})\,(v_{\rm drift}/v_{T})$, where $\eta_{O}^{\rm an} \sim |\eta_{H}|$ and $v_{\rm drift}/v_{T} \gtrsim 1$. So $\eta_{O} \sim |\eta_{H}|\,(v_{\rm drift}/v_{T}) \gtrsim |\eta_{H}|$ and the excited Ohmic term generically becomes larger than the Hall term. The outcome is that one cannot have a dominant Hall term {\em and} superthermal drift speeds \citep[see also][]{choueiri:1999.anomalous.resistivity.scaling.hall.parameter,shay:1999.whistler.mediated.reconnection.on.gyro.or.hall.timescale}. 

\subsection{Driving the System Towards Sub-Thermal Drift Speeds and Smooth Magnetic Gradients}
\label{sec:consequences:subthermal}

Consider what happens if superthermal drift arises, i.e.\ something induces gradients with $\ell_{B} \ll \ell_{\rm crit}$. As discussed in \S~\ref{sec:where}, if we only adopt the zero-drift coefficients $\eta_{O,H,A}^{0}$, then especially if the Hall term dominates, there is no guarantee that the non-ideal terms will actually reduce the drift speed (or increase $\ell_{B}$) because it is non-diffusive. But per \S~\ref{sec:consequences:order}, we see that with the corrected coefficients, $\eta_{H}$ does not dominate in this limit, so we are ensured that the nonideal terms behave diffusively (see Fig.~\ref{fig:etas}). This means that they will always act to smooth out the gradients $\ell_{B}$ until subthermal drift is restored. The timescale for this to occur will be $\sim \ell_{B}^{2}/{\rm MAX}[\eta_{O},\,\eta_{A}] \sim (\ell_{B}/v_{T})\,(v_{\rm drift}/v_{T})^{-2}\,\mathcal{F}_{\rm max}^{-1}$ where $v_{\rm drift}/v_{T} \gtrsim 1$ and $\mathcal{F}_{\rm max} \equiv {\rm MAX}[1,\, \Omega_{+}/\omega_{\rm coll},\, \omega_{p-}/\Omega_{-},\, \omega_{\rm coll}/\Omega_{-}] \gtrsim 1$. So generically this will occur on a timescale much shorter than the sound crossing time. 

So unless there are strongly supersonic extrinsic motions driving the ``ideal'' term ${\bf U} \times {\bf B}$ to compress $\ell_{B}$ -- essentially, strong (supersonic and super-\Alf{ic}) transverse shocks with a large energy source ``pumping'' the currents -- the corrected coefficients will generically ensure $\ell_{B}$ rapidly expands to restore subthermal drift ($\ell_{B} \gtrsim \ell_{\rm crit}$). Equivalently, strong magnetic field gradients are diffused away on scales $\lesssim \ell_{\rm crit}$, shown in Fig.~\ref{fig:lcrit}.

Note here that $\ell_{B} \equiv | \bar{\bf B} | / | \nabla \times \bar{\bf B} |$ is the global gradient length scale; so {\em weak} magnetic field gradients on scales $\lambda \ll \ell_{\rm crit}$ are still allowed. If we consider some small-amplitude structure with $ | \nabla \times \bar{\bf B} | \sim \delta B / \lambda$ and $\delta B \ll B$; then sub-thermal drift requires $\ell_{B} \sim \lambda \,|B|/|\delta B| \gtrsim \ell_{\rm crit}$. In this limit we can thus think of the physics above as setting an upper limit to the strength of field perturbations on scale $\lambda$, i.e.\ $ | \delta B|/|B| \lesssim \lambda / \ell_{\rm crit}$. This implies that the modification to MHD waves from the Hall effect will remain, but that these waves will experience non-linear damping above a certain (wavelength-dependent) amplitude owing to induced super-thermal drift enhancing the Ohmic resistivity. Recalling that $\ell_{\rm crit}^{\rm drift} \sim r_{g} / (f_{\rm ion} \beta^{\rm plasma})$, this becomes $|\delta B|/|B| \lesssim (\lambda / r_{g}) \, f_{\rm ion}\,\beta^{\rm plasma}$. We see that this places a serious constraint in very low $f_{\rm ion}$ plasmas for short-wavelength MHD waves and other perturbations. This is similar to effects known in high-$\beta^{\rm plasma}$ collisionless plasmas due to analogous instability-mediated ``anomalous viscosities'': \Alf\ waves cannot propagate \citep{squire:2017.kinetic.mhd.alfven} and sound waves are modified \citep{kunz:firehose} above a $\beta^{\rm plasma}$-dependent threshold.

\subsection{Behavior in the Zero-Charge Limit}
\label{sec:consequences:hydro}

A peculiar feature of the ``standard'' formulation of non-ideal MHD (Eq.~\ref{eqn:nonideal.classical}) is that it is {\em not} guaranteed to reduce to hydrodynamics as charge vanishes ($q_{i} n_{i} \rightarrow 0$). The total momentum equation (Eq.~\ref{eqn:mom.tot}) depends only on ${\bf J}\times {\bf B} \propto \nabla \times {\bf B} \times {\bf B}$ and is independent of $q_{i} n_{i}$, so the system can only reduce to hydrodynamics if these magnetic gradient terms vanish. If the Ohmic or ambipolar terms dominate over the Hall terms in the non-ideal diffusivities, they are diffusive and scale $\propto 1/q_{+} n_{+}$, so the diffusivity would become infinite and $\ell_{B}$ would increase, ensuring the fields rapidly escape and cannot couple strongly in the momentum equation, as desired. But recall the Hall term is non-diffusive and strong field gradients (i.e.\ strong forces on the neutrals) can persist indefinitely or even become sharper. And in the standard formulation, {\em whether} the Hall term dominates depends only on the ratio of the gyro to collision frequencies, independent of $q_{i} n_{i}$. So one can have arbitrarily strong MHD forces on neutrals sourced by a vanishingly small number of charged particles, which are implicitly assumed to be moving at infinitely fast speeds to carry the current. This is obviously unphysical.

Our proposed correction terms in Eq.~\ref{eqn:etas.general} restore physical behavior by ensuring that as $q_{i} n_{i} \rightarrow 0$ (which causes $v_{\rm drift} \rightarrow \infty$), the Ohmic term always dominates, causing the field to decay resistively. Thus, the magnetic diffusivity $\propto v_{\rm drift}/q_{i} n_{i} \propto 1/(q_{i}n_{i})^{2}$ becomes infinitely large more rapidly than any other term in the MHD equations.

\subsection{Maximum Magnetic Field Amplification}
\label{sec:consequences:amplification}

Imagine some collapsing gas wherein ${\bf B}$ is amplified via flux-freezing and/or some local dynamo, with global length scale $\sim R$. To maintain flux-freezing or have a dynamo on scales $\ll R$ obviously requires that the medium not be highly resistive, which requires (at a minimum) that the drifts remain subthermal, i.e.\ $\ell_{\rm crit}$ cannot exceed $R$. This sets an effective maximum $\bar{B}$, from Eq.~\ref{eqn:vdrift}: 
\begin{align}
\label{eqn:Bmax} \bar{B} \lesssim B_{\rm max} \sim \frac{4\pi e v_{T}}{c} n_{e} R \sim 0.2\, {\rm G} \left( \frac{T}{100 {\rm K}} \right)^{1/2} \left( \frac{n_{e} R}{\rm cm^{-3} au} \right)
\end{align}
(where we have used $\bar{n}_{-} |q_{-}| \sim e n_{e}$ for most regimes of interest; \S~\ref{sec:multispecies}). This has the interesting observationally testable consequence that $B_{\rm max} \propto v_{T} n_{e} R$, i.e.\ the maximum field is proportional to the free electron column/dispersion measure. For well-ionized systems ($n_{e} \sim n$) this is generally uninteresting as $B_{\rm max}$ is enormous (equivalently $\ell_{\rm crit}$ is tiny; \S~\ref{sec:where}), and if we assume spherical isothermal collapse ($n\propto R^{-3}$) $B_{\rm max} \propto n^{2/3}$ scales as steeply as isotropic spherical flux-freezing and/or supersonic dynamo amplification \citep{su:2016.weak.mhd.cond.visc.turbdiff.fx,su:fire.feedback.alters.magnetic.amplification.morphology}. But for weakly-ionized systems like circumstellar disks, Eq.~\ref{eqn:Bmax} is quite constraining. Moreover if we assume weakly-ionized ($n_{e} \ll n$) isothermal spherical collapse, with a constant ionization rate per neutral $\zeta$ and simple ion-electron recombination ($\dot{n}_{e} \sim n_{n} \zeta - \alpha_{\rm rec} n_{e} n_{i}$) so $n_{e} \propto n^{1/2}$, then we obtain $B_{\rm max} \propto n^{1/6}$. Note that for systems with a strong mean field, we might expect cylindrical collapse along an axis $R$ perpendicular to ${\bf B}$, so $n \propto R^{-2}$, which for the same ionization assumption gives $B_{\rm max} \propto n_{e} R \propto n^{1/2} R \propto R^{0} \propto n^{0}$ -- i.e.\ no amplification at all. In either case, $B_{\rm max}$ increases very weakly (far weaker than flux-freezing) with density.

\subsection{Is Non-Ideal Hall MHD Ever Important?}
\label{sec:consequences:hall}

From the above (\S~\ref{sec:consequences:order} \&\ Figs.~\ref{fig:regimes}-\ref{fig:etas}), the non-ideal Hall term can only strongly dominate in the induction equation when
$\Omega_{+} \ll \omega_{\rm coll} \ll \Omega_{-}$ and $|{\bf U}| \ll v_{\rm drift} \ll v_{T}$. 
This is not theoretically impossible; it is however challenging in practice to imagine in equilibrium on large scales. For example, it is only possible if $|{\bf U}|\ll v_{T,\,i}$, i.e.\ if we have some flow with negligible gas (sub-thermal) gas motions on some relevant global scale $\ell_{B}$, within some modest window of density and field strength (Fig.~\ref{fig:regimes}), with $\ell_{B} \gg \ell_{\rm crit}$.

Note that the Hall factor $(\nabla \times {\bf B}) \times{\bf B}$ is the same term that appears in the total momentum equation: so for the Hall term to be important in induction on large scales (e.g.\ to drive formation of current sheets or magnetic switches, or ``torques up'' the disk, as observed in the Hall-dominated regime using the zero-drift coefficients), it must necessarily appear in the momentum equation as a magnetic pressure or tension force. Further it is easy to verify that the interesting Hall terms generically correspond to compressible fluctuations ($\nabla \cdot [(\nabla \times {\bf B}) \times{\bf B}] \ne 0$). So we should compare both restoring magnetic-tension forces (with characteristic timescales $\sim \ell_{B}/v_{A}$) and pressure forces (with timescales $\sim \ell_{B}/v_{T}$), to the characteristic Hall induction timescale $\sim \ell_{B}^{2} / |\eta_{H}|$. Noting the scaling of $|\eta_{H}| \sim v_{\rm drift} \ell_{B}$, the Hall timescale is $\sim \ell_{B}/v_{\rm drift}$, so can only be shorter than thermal timescales if $v_{\rm drift} \gg v_{T}$, but in that superthermal limit the Ohmic term always becomes larger than Hall. So only rather contrived Hall-driven effects would be expected to persist.

Another way of saying this is to note $|\eta_{H}| \sim v_{\rm drift} \ell_{B} \sim v_{T} \ell_{\rm crit}$, so the effective propagation speed of Hall effects is $v_{H} \sim v_{T} \ell_{\rm crit}/\ell_{B}$ and is self-limiting to $v_{H} \lesssim v_{T}$. So there will always be faster speeds from e.g.\ restoring pressure forces, if those are involved, let alone if any other super-sonic speeds appear in the problem. In contrast, the diffusive Ohmic and ambipolar speeds ($v_{O}$ and $v_{A}$) are $v_{H}$ modified by additional factors of $\omega^{\rm coll}/\Omega$ (plus anomalous terms) or $\Omega/\omega^{\rm coll}$ respectively, so can be much larger than thermal speeds in principle.

For weak perturbations, as noted above (\S~\ref{sec:consequences:subthermal}), the Hall term could still be important. Consider an {\em incompressible} magnetic fluctuation which is also weak (amplitude $|\delta B| \ll (\lambda/\ell_{\rm crit})\,|B|$) and on sufficiently short wavelengths $\lambda \ll \ell_{\rm crit} \ll \ell_{B}$ such that we can boost to a frame moving with the mean ${\bf U}$ and have only small fluctuations $\delta {\bf U}(\lambda)$ on scale $\lambda$. If we are in the limit where the ``classical'' Hall term is larger than the Ohmic and ambipolar terms, then in this limit we will just recover the usual linearized Hall-MHD equations: i.e.\ that the Hall term modifies short-wavelength \Alf\ waves to ion-cyclotron and whistler waves. So our intuition for these is unchanged, except that when we consider finite-amplitude waves, their amplitude must remain very small with $|\delta B| \ll (\lambda/\ell_{\rm crit})\,|B|$  to avoid exciting current-driven instabilities.

So in short, while it is not strictly impossible to imagine a parameter space where the non-ideal, weakly-ionized Hall MHD term could be important for order-unity magnetic field changes/fluctuations, it is highly restricted. It seems likely in particular that many of the conclusions regarding e.g.\ thin current sheets in circumstellar disks (with short-axis width less than thermal pressure scale-lengths), or global ``disk torquing'' by the Hall effect, would need to be revised (Fig.~\ref{fig:lcrit}).

\subsection{Behavior in the Well-Ionized Limit}
\label{sec:ionized}

Briefly, consider the fully-ionized limit in some more detail. Returning to \S~\ref{sec:deriv}, if we have a fully-ionized electron-ion plasma we can quickly see $a_{O} \rightarrow \omega^{\rm coll}_{-+}/\Omega_{-}$, $a_{H} \rightarrow (1-\epsilon)/(1+\epsilon) \approx 1$, $a_{A}\rightarrow 0$. Thus the ambipolar term vanishes, and the Hall term is the largest non-ideal term in the ``classical'' ($v_{\rm drift} \ll v_{T}$) limit when $\Omega_{-} \gg \omega^{\rm coll}_{-+}$ (where $\omega^{\rm coll}_{-+}$ reflects e.g.\ the Spitzer/collisionless Coulomb interactions), which is easily satisfied in most low-density well-ionized astrophysical systems (except at extremely high densities and low temperatures). In this limit the term inside the induction equation becomes $\approx (\bar{\bf U} - {\bf v}_{\rm drift}) \times \bar{\bf B} \approx \bar{\bf u}_{e} \times \bar{\bf B}$, as expected.

In this limit, then, the Hall term can become important on some (small) scale $\lambda$ so long as $|\delta {\bf U}(\lambda)| \lesssim v_{\rm drift} \lesssim v_{T}$. In the fully ionized limit the drift velocities are generally small, $v_{\rm drift} \sim v_{T} \ell_{\rm crit}/\ell_{B} \sim (v_{T}/\beta^{\rm plasma})\, (r_{g} / \ell_{B}) \sim (v_{T}/\beta^{\rm plasma})\, (r_{g} / \lambda)\, (|\delta B|/|B|)$. Thus the Hall term being important requires $|\delta {\bf U}(\lambda)|/v_{T} \ll (r_{g} / \beta^{\rm plasma} \lambda)\,(|\delta B|/|B|) \ll 1$. Since MHD is only valid on scales $\lambda \gg r_{g}$, the latter inequality ($v_{\rm drift} \ll v_{T}$ being sub-thermal) is essentially always satisfied, while the former inequality ($|\delta {\bf U}(\lambda)| \ll v_{\rm drift}$) is valid on sufficiently small scales $\lambda$. Note that on large scales in ISM and other applications, $|\delta {\bf U}(\lambda)|/v_{T} \gtrsim 1$, i.e.\ bulk motions are trans or super-sonic, and $\lambda \gg r_{g}$ by a huge factor while $|\delta B| \lesssim |B|$, which means that the Hall term is completely negligible compared to the ideal MHD term ${\bf U} \times {\bf B}$. This is just a restatement of the fact that ideal MHD is an excellent approximation on large scales in well-ionized astrophysical plasmas. But this again leads to the usual expected small-scale behaviors that Hall terms will modify short-wavelength \Alf\ waves as expected in a well-ionized plasma, though other effects (e.g. electron inertia) can also become important on small scales \citep{schekochihin:2009.gyrokinetics.cascades.weakly.collisional.plasmas}.

\section{Conclusions}
\label{sec:conclusions}

The commonly-adopted astrophysical MHD equations are no longer internally self-consistent if gradients in ${\bf B}$ become too steep (gradient scale-length $\ell_{B}$ smaller than some critical $\ell_{\rm crit}$), because the implied drift speeds ($v_{\rm drift}$) become much faster than thermal speeds ($v_{T}$). This can invalidate foundational assumptions of the equations, rendering the expressions generally adopted for certain terms (like the non-ideal Ohmic/Hall/ambipolar coefficients) incorrect. Most importantly, it will excite microscale instabilities that suppress the drift, potentially modifying the effective collision frequencies by orders of magnitude. We show that this could be relevant in weakly ionized systems like cold cores, circumstellar/planetary disks, or planetary atmospheres, especially if (1) the Hall effect is significant, (2) external forces (like gravity) induce highly supersonic motions, or (3) dust grains become an important charge carrier. The ``classical'' treatments could then predict unphysical behaviors. We derive a simple prescription to remedy this, extending treatments established in the fusion, Solar, and ionospheric plasma literature, and directly analogous to well-established treatments for similar problems in thermal conduction and viscosity. The treatment amounts to simply multiplying the Ohmic and ambipolar coefficients by factors involving only macroscopic (already-calculated) quantities. Implementing these correction terms in simulations therefore entails negligible cost or complexity. They also leave the behavior unchanged in the limit of slow drift for which the classical expressions are derived, but will restore the physical behavior if this limit is violated, driving the system back towards slow drift. 

We show that this leads to several important behaviors in weakly-ionized systems. First, the Hall term can never dominate among non-ideal (weakly-ionized) terms if the drifts are superthermal, which ensures the behavior of these terms is diffusive. Second, this rapidly drives the system back to subthermal drifts by diffusing away any strong gradients in the magnetic field on length scales smaller than $\ell_{\rm crit}$, which can be quite large (larger than protostellar disk sizes). Third, this ensures intuitive behaviors such as the system becoming hydrodynamic if the ionization fraction vanishes, which is not strictly ensured by the classical equations. Fourth, this can strongly restrict the maximum magnetic-field amplification during collapse such that it scales well below the flux-freezing estimate. And fifth, we show that under these restrictions, it may be difficult (if not impossible) for weakly-ionized systems to ever achieve quasi-steady-state conditions where the Hall term could be dynamically important on large scales, as we show these generically require superthermal drift speeds. 

In future work, it will be important to study the practical effects of Eq.~\ref{eqn:nonideal.eff} in numerical simulations of weakly-ionized media. But of course this will be problem-specific and the effects could, by construction, vary from negligible to complete magnetic decoupling. Our hope here is that the simple prescription provided can inform future simulations of weakly ionized systems, being used as a rough approximation to maintain consistency and probe both where the microphysics should be important and how it should limit macroscopic consequences. Of course, it will also be valuable to explore PIC or MHD-PIC type simulations of these behaviors, especially in the particularly ambiguous dust-charge-dominated limits (see \citealt{ji:2021.cr.mhd.pic.dust.sims,ji:2021.mhd.pic.rsol}). Such studies are needed better inform the precise scalings of $\eta_{\rm an}$ especially in more extreme regimes.

\begin{acknowledgements}
Support for PFH was provided by NSF Research Grants 1911233, 20009234, 2108318, NSF CAREER grant 1455342, NASA grants 80NSSC18K0562, HST-AR-15800. JS acknowledges the support of the Royal Society Te Ap\=arangi, through Marsden-Fund grant  MFP-UOO2221 and Rutherford Discovery Fellowship  RDF-U001804.
\end{acknowledgements}

\bibliographystyle{mn2e}
\bibliography{ms_extracted}

\begin{appendix}

\section{Additional Multi-Species Details, Numerical Solutions, and Comparisons}
\label{sec:extra.multi}

In the main text, for the arbitrary multi-species case (\S~\ref{sec:multispecies}), we propose a simple weighting of the effective Hall/Ohmic/ambipolar coefficients to capture the leading-order anomalous resistivity effects, together with the approximate analytic rescalings in Eq.~\ref{eqn:etas.general.specific}. Here we discuss the standard multi-species derivation of the non-ideal terms in more detail, and calculate more accurate solutions numerically, to show that our proposed corrections capture the relevant behavior in the limits of interest.

\subsection{Overview and Classical Derivation}
\label{sec:extra.multi:ov}

Consider the case with an arbitrary number $N$ of charged species. In order to obtain any closed set of equations on macro scales, impose from \S~\ref{sec:deriv.nonideal.mhd} assumptions {\bf (1)} (Ampere's law), {\bf (2)} (drop $D_{t}$, stress, and battery terms), and {\bf (5)} (neglect micro/meso fluctuations), and assume one dominant neutral species (for multiple neutrals some other equations must be introduced to close their relative slip velocities). The momentum equation for each charged species $j$ then reduces to ${\bf E} \approx (B/c) \bhat \times {\bf u}_{j} + (1/n_{j} q_{j})\sum_{i}[ \rho_{j} \omega_{ji} ({\bf u}_{i} - {\bf u}_{j}) - \dot{\rho}_{ji} {\bf u}_{j} + \dot{\rho}_{ij} {\bf u}_{i} ]$. Define the $3N$-element vectors ${\bf V} \equiv \{ u_{0}^{x},\, u_{0}^{y},\, u_{0}^{z},\, u_{1}^{x},\,u_{1}^{x},\,u_{1}^{z},\,u_{2}^{x}\,...\}$ ($V_{n} = u_i^{\alpha}$ with $i={\lfloor n/3 \rfloor}$ the species and $\alpha = n-{\lfloor n/3 \rfloor}$ the directional component) and $\tilde{\bf E} \equiv \{ E^{x},\,E^{y},\,E^{z},\,E^{x},\,... \}$ ($\tilde{E}_{n} = E_{\alpha}$). Then we can write the momentum equation $\tilde{\bf E} = {\bf S} {\bf V}$ where ${\bf S}$ is $3N\times 3N$ and invertible, so ${\bf V} = {\bf S}^{-1} \tilde{\bf E}$. Because of the repetitive nature of $\tilde{\bf E}$ we can decompose ${\bf S}^{-1} = {\bf S}^{\prime}_{{\rm L}} {\bf S}^{\prime}_{{\rm R}}$ into the $3N\times 3$ and $3 \times 3N$ left/right blocks ${\bf S}^{\prime}_{{\rm L}}$ and ${\bf S}^{\prime}_{{\rm R}}$, respectively, with ${\bf S}^{\prime}_{{\rm R}} \cdot \tilde{\bf E} = {\bf E}$, so ${\bf V} = {\bf S}^{\prime}_{{\rm L}} \cdot {\bf E}$. Then we impose $\sum_{j} n_{j} q_{j} {\bf u}_{j} = {\bf J} = {\bf J}_{A}$, which we can write ${\bf A}_{{\rm R}} {\bf V} = {\bf J}_{A}$ where ${\bf A}_{{\rm R}}$ is $3 \times 3N$, giving $({\bf A}_{\rm R} {\bf S}^{\prime}_{\rm L}) {\bf E} = {\bf J}_{A}$, where $({\bf A}_{\rm R} {\bf S}^{\prime}_{\rm L})$ is a $3\times3$ invertible matrix (even though ${\bf A}_{\rm R}$ and ${\bf S}^{\prime}_{\rm L}$ are not invertible). So we obtain the solutions for both ${\bf E}$ and the various ${\bf u}_{j}$ as ${\bf E} = ({\bf A}_{\rm R} {\bf S}^{\prime}_{\rm L})^{-1} \cdot {\bf J}_{A}$ and ${\bf V} = {\bf S}^{\prime}_{\rm L}  ({\bf A}_{\rm R} {\bf S}^{\prime}_{\rm L})^{-1} \cdot {\bf J}_{A}$. Finally noting that we can write the basis vector $\hat{\bf J}^{\prime} \equiv \{ \hat{\bf J} ,\, \hat{\bf J} \times \bhat,\, \hat{\bf J}\times\bhat \times \bhat \} = {\bf R}\, \hat{\bf x}$, we can express ${\bf E}$ in terms of the components $\hat{\bf J}^{\prime}$ as ${\bf E}^{\prime} \rightarrow ({\bf R}^{-1})^{T} {\bf E}$ to write the MHD induction equation in terms of some effective Ohmic, Hall, and ambipolar coefficients\footnote{If we locally define coordinate axes such that $\bhat=\{0,0,1\}$ and ${\bf J} = |{\bf J}| \{\sin{\theta},\,0,\,\cos{\theta}\}$, then $\eta_{O} = (c^{2}/4\pi |{\bf J}|)\,E_{z}/\cos{\theta}$, $\eta_{H}=-(c^{2}/4\pi |{\bf J}|) E_{y}/\sin{\theta}$, $\eta_{A} = (c^{2}/4\pi |{\bf J}|) E_{x}/\sin{\theta} - \eta_{O}$.} $\eta_{O,H,A}$. 

In principle, one could take this approach and simply use the drift/slip-velocity-dependent form of all the collision rates: $\omega_{ij} \rightarrow \omega_{ij}({\bf V},\,...) + \omega^{\rm an}_{ij}({\bf V},\,...)$. 
This accounts for both the classical dependence of the direct collision rates $\omega_{ij}({\bf V},\,...)$ on all the relative velocities ${\bf u}_{j}$ (in ${\bf V}$; e.g.\ \citealt{pinto.galli:2008.momentum.transfer.coefficients.for.weakly.ionized.systems}) as well as the anomalous resistivity/collision rate of each species with respect to others (added as an effective collisionality per \S~\ref{sec:deriv.more}, itself also a function of ${\bf V}$ through e.g.\ functions like $\Theta({\bf u}_{j}-{\bf u}_{i})$ in Eq.~\ref{eqn:etas.general.specific}). 
Similarly one could modify the inelastic and charge exchange rates $\dot{\rho}$ to be appropriate functions of ${\bf V}$ as well. 
But then the terms appearing in ${\bf S}$ are highly non-linear functions of the different components of ${\bf V}$ and (implicitly) ${\bf E}$ and ${\bf J}$, and multi-dimensional iterative root-finding methods are needed. And even if we took the ``zero drift/slip'' coefficients ($\omega_{ij}({\bf V}) \rightarrow \omega_{ij}({\bf V}=\boldsymbol{0})$) so ${\bf S}$ were independent of ${\bf V}$, if we retain general cross-interaction terms $\omega_{ij}$ and charge-exchange/inelastic terms then the matrix solve remains quite complicated. Still, for methods that actually numerically solve a chemical network on-the-fly for the abundances of $N$ directly, this is not necessarily more expensive than the network solve itself, so may be preferred. 

However, most implementations either pre-compute chemical tables or otherwise simplify to only use some end product of this series of inversions. Moreover, even if we followed the non-linear procedure above, for general $\omega_{ij}({\bf V})$ and $N$, there is no guarantee of a {\em unique} solution ${\bf V}$ for given ${\bf J}$ (there can be multiple viable sets of ${\bf u}_{j}$, which would require evolving $D_{t}{\bf u}_{j}$ to distinguish, depending on how the $\omega_{ij}$ terms depend on ${\bf V}$). 

We can now see the primary motivation for the common literature assumptions {\bf (3)} (assume dominant neutrals with ${\bf U} \approx {\bf u}_{n}$, and neglect all $\dot{\rho}_{ij}$ and non-neutral collision $\omega_{ij}$ terms) and {\bf (4)} (assume vanishing drift/slip so $\omega_{ij} \rightarrow \omega_{ij}({\bf V}=\boldsymbol{0})$). With these simplifications, ${\bf S}$ has a trivial block form and the equation for each species is separable: ${\bf W} \equiv (c/B)\,({\bf E} + {\bf U}\times{\bf B}/c) \approx \bhat \times \delta{\bf u}_{jn} + \tilde{\beta}_{j}^{-1} \delta {\bf u}_{jn}$ (where $\delta {\bf u}_{jn} \equiv {\bf u}_{j} - {\bf u}_{n}$ are the {\em slip} velocities and $\tilde{\beta}_{j} \equiv q_{j} B/m_{j} c \omega_{jn}$). This gives the familiar result that the relative contributions to the current for each species depend only on $\tilde{\beta}_{j}$, and coefficients $\eta_{O} = (c B/4\pi) \tilde{\sigma}_{O}^{-1}$, $\eta_{H}=(c B/4\pi) \tilde{\sigma}_{H}/\tilde{\sigma}_{\bot}^{2}$, $\eta_{A}=(c B/4\pi) (\tilde{\sigma}_{P}/\tilde{\sigma}_{\bot}^{2}-1/\tilde{\sigma}_{O})$ where $\tilde{\sigma}_{O} \equiv \sum n_{j} q_{j} \tilde{\beta}_{j}$, $\tilde{\sigma}_{H} \equiv n_{j} q_{j}  /(1+\tilde{\beta}_{j}^{2})$, $\tilde{\sigma}_{P} \equiv n_{j} q_{j} \tilde{\beta}_{j}/(1+\tilde{\beta}_{j}^{2})$, $\tilde{\sigma}_{\bot}^{2} \equiv \tilde{\sigma}_{H}^{2} + \tilde{\sigma}_{P}^{2}$. With the slip velocities\footnote{The slip velocities are given in a frame where 
$\bhat=\{0,0,1\} = \hat{z}$ by: 
\begin{align}
\nonumber \delta{\bf u}_{jn,x} &= 
\frac{\tilde{\beta}_{j} (W_{x} + \tilde{\beta}_{j} W_{y})}{(1+\tilde{\beta}_{j}^{2})}
= 
 \frac{\tilde{\beta}_{j} [J_{x}(\tilde{\sigma}_{P} - \tilde{\beta}_{j} \tilde{\sigma}_{H} ) + J_{y}(\tilde{\sigma}_{H} + \tilde{\beta}_{j} \tilde{\sigma}_{P} )]}{(1+\tilde{\beta}_{j}^{2})\sigma_{\bot}^{2}}  
 \ , \\
\nonumber \delta{\bf u}_{jn,y} &= 
\frac{\tilde{\beta}_{j} (W_{y} -\tilde{\beta}_{j}  W_{x} )}{(1+\tilde{\beta}_{j}^{2})}
= 
\frac{\tilde{\beta}_{j} [ J_{y}(\tilde{\sigma}_{P} - \tilde{\beta}_{j} \tilde{\sigma}_{H} ) -J_{x}(\tilde{\sigma}_{H} + \tilde{\beta}_{j} \tilde{\sigma}_{P} )]}{(1+\tilde{\beta}_{j}^{2})\sigma_{\bot}^{2}}
\ , \\
\delta{\bf u}_{jn,z} &= 
\tilde{\beta}_{j} W_{z} 
= 
 \frac{\tilde{\beta}_{j} J_{z}}{\tilde{\sigma}_{O}} \ .
\end{align}
} $\delta {\bf u}_{j}$, we can calculate each pairwise drift ($\delta {\bf u}_{j} - \delta {\bf u}_{i}$), and again in principle modify $\omega_{ij}$ to account for each drift+slip speed. But this again introduces the difficulties above: we then modify $\tilde{\beta}_{j}$ so must iteratively re-calculate the relative drift/slip speeds, are not ensured a unique solution, and (conceptually) must evaluate ``which drifts matter.''\footnote{For example, even if electrons carry most of the current, since we only retain their effective collision rate with neutrals here, do we include anomalous resistivity if their drift is subthermal relative to a dominant ion in the current but superthermal relative to a more trace ion, like charged grains?}

\begin{figure*}
	\centering\includegraphics[width=0.95\columnwidth]{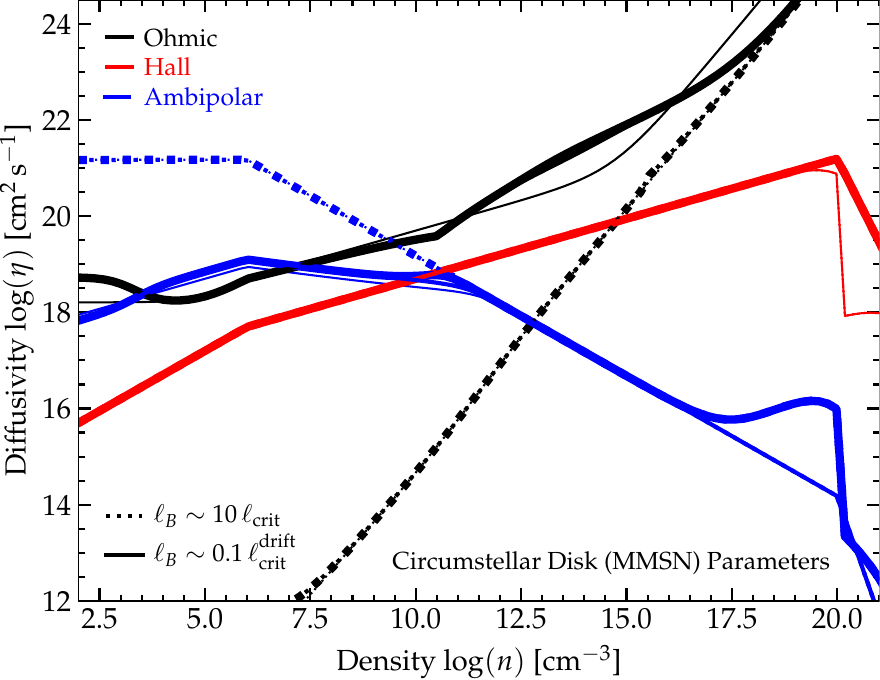} 
	\centering\includegraphics[width=0.95\columnwidth]{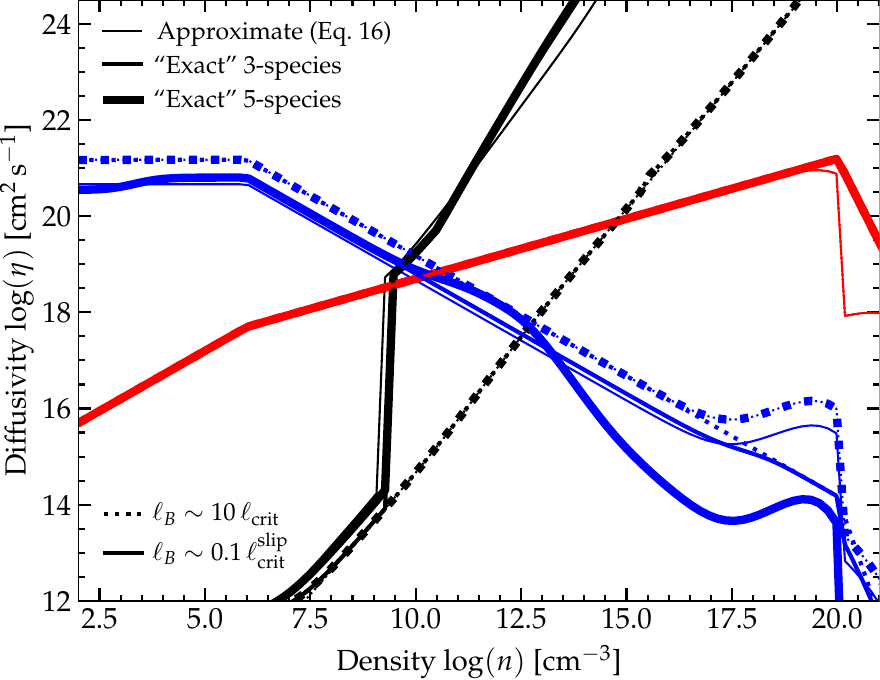} 
	\caption{Ohmic, Hall, and ambipolar diffusivities $\eta_{O,H,A}$ as a function of density as in Fig.~\ref{fig:etas} (for the ``circumstellar disk''-like case), but for the specific numerical examples in \S~\ref{sec:numerical.example}.
	{\em Left:} Comparison of a generally-subthermal case ($\ell_{B} = 10\,\ell_{\rm crit}$; {\em dotted}) to a superthermal drift case ($\ell_{B} = 0.1\,\ell_{\rm crit}^{\rm drift}$; {\em solid}). 
	{\em Right:} Same but for the superthermal slip case $\ell_{B}=0.1\,\ell_{\rm crit}^{\rm slip}$.
	For each, we compare 
	(1) the prediction from the simple parameterized rescaling we propose in Eq.~\ref{eqn:etas.general.specific} (for the multi-species scalings in \S~\ref{sec:multispecies}; ``Approximate'' in {\em thin} lines); to 
	(2) the case where ignore charged dust grains as a current carrier (treat them as part of the neutral inertia taking $\tilde{\beta}_{\rm grain} \rightarrow 0$) and solve the full non-linear three-species problem with the exact coefficients (\S~\ref{sec:deriv}; ``Exact'' 3-species in {\em intermediate} lines); and 
	(2) the results of exactly solving the non-linear multi-species equations as described in \S~\ref{sec:extra.multi:ov} with all possible pairs of species featuring simple Epstein-type scalings and anomalous terms (\S~\ref{sec:numerical.example}; ``Exact'' 5-species in {\em thick} lines). 
	The 3 and 5-species methods give excellent agreement, indicating that grain current and differences between exact velocity-dependent scalings in the superthermal regime make little difference. The closed-form analytic approximate scaling from Eq.~\ref{eqn:etas.general.specific} reproduces these reasonably well in all regimes of interest. Key qualitative results (e.g.\ vanishing of the Hall-dominated regime) are robust.
	\label{fig:etas.numerical}}
\end{figure*}

\subsection{Application to Scalings In the Text}

Here we comment on how well the approximate multi-species scalings in the text \S~\ref{sec:multispecies} capture the previous discussion.
Despite the challenges of solving the full system in generality with superthermal drifts and anomalous terms, it is nonetheless possible to make some robust statements. 

First, the leading-order contribution to the charged drift speeds $|{\bf u}_{i}-{\bf u}_{j}|$ always scales as $\propto \tilde{\beta}_{i} - \tilde{\beta}_{j}$ (whether the various $\tilde{\beta}_{i,j}$ are $\ll 1$ or $\gg 1$), and therefore contributions to the current scale $\propto \tilde{\beta}_{j}$ as we noted in \S~\ref{sec:multispecies}. More quantitatively, one can show that our approximate expression for the ``effective'' drift $v_{\rm drift}$ (Eq.~\ref{eqn:nq}) and slip (Eq.~\ref{eqn:vdrift}) $v_{\rm slip,\,max}$ velocities in the text is exact (for the dominant current-carrying species) in the limits of either (a) two dominant species containing most charge and current (one positive one negative); (b) one dominant species in the current (one species with $|\tilde{\beta}_{j}|$ or $|n_{i} q_{i} {\bf u}_{i}|$ much larger than any other of the salient charge-carriers); or (c) all salient $|\tilde{\beta}_{j}|$ small. And even in the worst-case scenario (all $|\tilde{\beta}_{j}|$ large with several species carrying comparable current) these expressions still approximate the mean drift/slip velocity to within a factor of $\sim 2$. Given the ambiguity in these cases and our approach of even defining ``which'' $v_{\rm drift}$ matters and the order-unity ambiguity in defining the threshold drift for anomalous resistivity (depending on precisely which instabilities are active and how), this uncertainty is more than adequate for our purposes. 

We can also see that if we multiply all collision rates by some factor $\psi$, then $\eta_{O}\rightarrow \eta_{O}\,\psi$, while for all salient limits (when the various dominant charge-carrier $|\tilde{\beta}_{j}|$ are $\ll 1$ [Ohmic], or $\gg 1$ [ambipolar], or some are $\ll 1$ and some $\gg 1$ [Hall], or when there is a single dominant carrier), $\eta_{A} \rightarrow \eta_{A}/\psi$, while $\eta_{H}$ is unmodified. Again even in the worst-case scenario (many nearly-equal $\tilde{\beta}_{i}$), this at worst equates to an $\mathcal{O}(1)$ uniform difference in the pre-factor of $\eta_{H}$. This justifies the correction in the text for the Epstein drag limit (see also \citealt{hillier:2024.ambipolar.drift.w.supersonic.vdrift}, who consider Epstein drag corrections with a more restricted set of assumptions for just ambipolar drag in the two-fluid limit). Finally, if we return to the original equation and consider adding a constant $\omega^{\rm an}$ which represents a collision frequency between {\em charge carriers} (either between them and each-other, or some comoving \Alf\ wavepacket, or similar concept), we can see that in the limits above or any limit where the Ohmic term would dominate ($|\tilde{\beta}| \ll 1$), in $\sigma_{O}$ we should simply take $\tilde{\beta}_{j} \rightarrow q_{j} B/m_{j} c \omega^{\rm tot}_{j}$ where $\omega^{\rm tot}_{j} \equiv \sum_{i} \omega_{ji}$ includes {\em all} collision terms influencing $j$. This is explicit in our three-species derivation in the text, where $\omega^{\rm tot}$ includes terms like $\omega_{-+}$ between charge-carriers. This means $\eta_{O} \propto \omega^{\rm tot} \sim \omega^{\rm classical} + \omega^{\rm an}$ for whatever dominant species $j$ contributes the most in $\sigma_{O}$, i.e.\ that we can justify the addition of $\omega^{\rm an}$ as in Eq.~\ref{eqn:etas.general} in the text. 

In contrast, we find for any limit where the ambipolar term is dominant in the classical limit ($\tilde{\beta} \gg 1$ ignoring $\omega^{\rm an}$), that {\em only} collisions with neutrals contribute to leading order in $\eta_{A}$. Again, this is explicit in our three-species derivation (only $\omega_{\pm n}$ terms appear), and obvious in the sense that the ambipolar limit physically represents neutral-charge carrier slip independent of the charge-carrier drift velocities amongst themselves. As noted in the text, there certainly are instabilities that grow when the slip becomes superthermal, but their non-linear outcomes are less clear. Nonetheless one could add them if desired to the equivalent $\beta$, if desired. For the reasons discussed above, the Hall term is again unmodified in either limit by these additions, and even in the worst-case scenario is only modified by an $\mathcal{O}(1)$ prefactor, as in our three-species derivation.

\subsection{Numerical Examples}
\label{sec:numerical.example}

To numerically validate the above and our approach, we revisit our simple example in the text of Fig.~\ref{fig:etas}, where we calculate the $\eta_{O,H,A}$ coefficients according to some chemical model, but solve the generalized non-linear multidimensional problem exactly as described above (\S~\ref{sec:extra.multi:ov}; making assumptions {\bf (1)}, {\bf (2)}, {\bf (5)} to close the system, but no others). 
This cannot be done in NICIL without extensive code modifications, so we instead adopt an extremely simple chemistry model for the quantities needed to compute the values shown. Specifically, we begin from (with all units here in CGS) similar parameters to those in the NICIL default model in-text \citep{wurster:2021.do.we.need.nonideal.mhd} with temperature $T=10 \sqrt{1+(n/10^{n_{T,1}})^{0.8}} (1+n/10^{n_{T,2}})^{-0.3} (1+n/10^{n_{T,3}})^{0.57}$ for $(n_{T,1},n_{T,2},n_{T,3})=(11,16,21)$ from \citet{machida:2006.second.core.form.simplified.eos.sims}; and $B=10^{-6} n^{1/2}$ for $n<n_{0,B}$, $B=10^{-6} (n n_{0,B})^{1/4}$ for $n\ge n_{0,B}$ with $n_{0,B}=10^{6}$ from \citet{wardle.ng:1999.nonideal.coefficients}. We adopt a simple five-species model of: molecular neutrals with Solar abundances ($\xi_{n} \approx 1$; $m_{n}=2.3 m_{p}$); electrons ($n_{e} \sim 0.01$ for $n<10^{n_{e,1}}$ and $n_{e} \sim 0.01 (n/10^{n_{e,1}})^{2}$ for $n\ge10^{n_{e,1}}$, with $n_{e,1}=14$); atomic+molecular ions ($n_{i} = n_{e}$, $q_{i}=e$, $\langle m_{i} \rangle=29 m_{p}$); negative and positive small dust grains ($n_{g+}=n_{g-} \sim 10^{-13} n$, $q_{g+}=-q_{g-}=e$, $R_{\rm grain}=3 \times10^{-6}$). To be more appropriate for a circumstellar disk, we adjust the model parameters above to give  similar results to the minimum mass Solar nebula (MMSN) model from \citet{chiang:2010.planetesimal.formation.review}, taking $(n_{T,1},n_{T,2},n_{T,3}) \rightarrow (10.5,12.5,19.5)$, $n_{e,1}\rightarrow 20$, and $n_{g\pm} \rightarrow 10^{-15} n$ at $n<10^{n_{e,1}}$.
We take the zero-drift cross sections to have values $\langle \sigma v \rangle_{in}=1.5 \times10^{-9}$, $\langle \sigma v \rangle_{en}=4.7 \times10^{-10} T^{1/2}$, $\langle \sigma v \rangle_{gn}=5.1 \times10^{-6} T^{1/2}$, $\langle \sigma v \rangle_{ei}=51 T^{-3/2}$ \citep{spitzer:conductivity} and neglect any other reactions. 

We now have interactions between electrons, heavy ions, neutrals, positively and negatively charged grains. Per \S~\ref{sec:extra.multi:ov}, to even write down the equations for ${\bf E}$ in this limit we must define the velocity-dependent interaction terms between all pairs of particles, each of which can have different instability thresholds and qualitative behaviors of their anomalous terms (many of which remain poorly understood in detail). So we necessarily adopt a simple toy model, with the purpose not of providing any sort of ``first principles'' calculation, but simply to validate the simple approximate scalings discussed above. We neglect charge-exchange reactions, but consider all effective collisional cross terms. This includes both a ``classical'' (but correctly velocity-dependent) rate which we approximate with a simple hard-sphere-scattering type expression $\omega^{\rm coll}_{ji} = \omega^{\rm coll}_{ji}(\delta {\bf u}_{ji}\rightarrow 0) \times \left[ 1 + |\delta {\bf u}_{ji}|^{2}/v_{T,\,ij}^{2}\right]^{1/2}$ (for which we use the values of $\langle \sigma v (\delta {\bf u}_{ji}\rightarrow 0) \rangle$ in text, supplemented with the zero-drift ion-grain and electron-grain cross sections from \citealt{draine:1987.grain.charging}), plus a pairwise anomalous term to capture different instabilities that can occur between each pair of interacting charged species, $\omega^{\rm an}_{ji} = {\rm MAX}[\Omega_{j},\,\Omega_{i},\,\omega_{pj},\,\omega_{pi}]\,\Theta(|\delta {\bf u}_{ji}|/v_{T,ij})$ for $j,\,i\ne n$, with $v_{T,\,ij} = {\rm MIN}[v_{T,\,i},\,v_{T,\,j}]$ (per Eqs.~\ref{eqn:vT}-\ref{eqn:vA.eff}, with $\psi_{j}=0$ for $j\ne i,n$, so just the effective speeds of the individual carriers plus potentially-entrained neutrals for each). Like in Fig.~\ref{fig:etas} we consider two limits, which we define by the magnetic-field gradient $\ell_{B} = 10\,\ell_{\rm crit}$ and $\ell_{B} = 0.1\,\ell_{\rm crit}^{\rm drift}$ or $\ell_{B} = 0.1\,\ell_{\rm crit}^{\rm slip}$ (with $v_{T} = {\rm MIN}[v_{T,\,n},\,v_{A,\,{\rm ideal}}]$ in $\ell_{\rm crit}$). In the model in the text, these correspond one-to-one to the sub-thermal ($v_{\rm drift} \sim 0.1\,v_{T}$) and super-thermal ($v_{\rm drift}\sim 10\,v_{T}$) limits, but since there is no single drift velocity to speak of in the more general case, they  must be defined in terms of $\ell_{B}$ (which is independent of the collision rates and species involved). Because of the non-linear dependencies involved, we must also specify the geometry, namely $\cos{\theta} \equiv \hat{\bf J} \cdot \bhat$, for which we take a representative $\theta=30^{\circ}$, but note the results are very weakly dependent on $\theta$, except if very close to parallel ($|\cos{\theta}|\approx 1$) where the ambipolar and Hall terms trivially vanish. Again, we emphasize the various coefficients and rates above are not meant to be exact, but simply to provide some calculable illustration.

We further consider one more comparison. If we treat the dust grains as simply a ``charge sink'' and assume their mass to be anchored to the neutrals (adding to neutral inertia to leading order), then our system reduces to a three-component (neutrals, ions, electrons) system. We take this, plus the full velocity-dependent scattering rates from \citet{pinto.galli:2008.momentum.transfer.coefficients.for.weakly.ionized.systems} for each component, with a single ion-electron anomalous resistivity added to the ion-electron collision rate (where present), and otherwise adopt the scalings above. Then we can apply our solution from the main text (\S~\ref{sec:deriv}) and exactly solve the non-linear problem with all terms from \S~\ref{sec:deriv} included (still making assumptions {\bf (1)}, {\bf (2)}, {\bf (5)} required for closure). 

In Fig.~\ref{fig:etas.numerical} we see that our extremely simplified proposed rescaling of the zero-drift coefficients (Eq.~\ref{eqn:etas.general.specific}) provides a reasonable approximation to either of these more complicated calculations (both of which agree very well with each other), for both large and small $\ell_{B}/\ell_{\rm crit}$, at all densities plotted. The most important reason for this is that for all the regimes of interest here, even when grains contain most of the charge, the current is always primarily being carried by the free electrons. In the limit of a single dominant current carrier, all that matters to leading order is the effective collision rate for that species, whose drift velocity must be set by $\ell_{B}$ (nearly independently of the collision rates) if Ampere's law is to hold.  The similarity between the two more complicated numerical cases (the five- and three-species models) further emphasizes that the dust current is not important. 

The residual deviations between the simple analytic approximation and the more complicated calculations at superthermal drift speeds primarily owe to subtleties of where the various drift and slip terms appear. For example, at high-$n$, the drift and slip corrections to $\eta_{A}$ cannot be trivially combined into a single effective coefficient as we do in Eq.~\ref{eqn:etas.general.specific}, and this leads to the systematic offset between Eq.~\ref{eqn:etas.general.specific} and the more general solutions. Similarly, dust effects play some role in $\eta_{A}$ at the highest densities, and $\eta_{H}$ has offsetting terms in the velocity-dependent case that make its ``jump'' at $n\gtrsim 10^{20}\,{\rm cm^{-3}}$ more smooth. But all of these differences are in the regimes where those terms are much smaller than the Ohmic term, so do not dynamically matter. For the Ohmic term the deviations at high densities owe to a mix of the above effects, as well as the fact that when the anomalous  and classical resistivities are comparable, they add nonlinearly in the more exact formalisms (like that in \S~\ref{sec:deriv} and as implemented in the 3 or 5-species models).

\section{Analytic Approximations for Modified Coefficients in the Superthermal Limit}
\label{sec:deriv.more}

\subsection{Modified ``Standard'' Collision Rates}
\label{sec:epstein.deriv}

Here we briefly present a more detailed justification from our three-species model in \S~\ref{sec:deriv} of the approximate scalings for $\eta_{O,\,H,\,A}$ that appear in the superthermal limit owing to modified (velocity-dependent) collision rates (ignoring anomalous terms for now).

Let us approximate the collision rates with the Epstein scalings (as applicable to neutral hard-sphere collisions, technically) for sufficiently large $\Delta u_{ij}$ (the case of interest), $\omega^{\rm coll}_{ij} \rightarrow \omega^{0}_{ij}\,(1 + \Delta u_{ij}/v_{T,\,\rm eff}^{2})^{1/2}$ (where $v_{T,\,{\rm eff}} = (8/3\pi^{1/2}) v_{T,\,ij}$ and $\Delta u_{ij} \equiv |{\bf u}_{i}-{\bf u}_{j}|$), but retain all other assumptions from \S~\ref{sec:deriv.nonideal.mhd}. Then it is straightforward to see that Eq.~\ref{eqn:nonideal.classical} and Eq.~\ref{eqn:vdrift} still apply, but the coefficients $\omega_{O,H,A}$ must use their correct (self-consistent) values for the given $\Delta u_{ij}$, which in turn modifies $\eta_{O,H,A} \propto a_{O,H,A} \propto \omega_{O,H,A}$. But none of the other parameters (e.g.\ the various $\Omega$, or $\epsilon$, $\xi$, $B$, $c$, $Q$, ${\bf J}_{A}$ terms) in Eqs.~\ref{eqn:Ohms}-\ref{eqn:vdrift} are modified. That means ${\bf v}_{\rm drift}$ is unmodified, but ${\bf v}_{\rm slip}^{\pm}$ depend on the various $\omega_{in}$ so terms like $\Delta u_{in}$ are defined implicitly in Eq.~\ref{eqn:vdrift}. 
From Eq.~\ref{eqn:Ohms}, we immediately see that if we increase all $\omega_{ij}$ by the same factor $f_{v}$, then $\omega_{O} \rightarrow \omega_{O}^{0}\,f_{v}$, $\omega_{H} \rightarrow \omega_{H}^{0}$, $\omega_{A} \rightarrow \omega_{A}^{0}/f_{v}$. But $f_{v} = \sqrt{1 + \Delta u_{ij}^{2}/v_{T,\,\rm eff}}$ depends on species. 

For an arbitrary set of species, we can modify this species-by-species and derive the ``exact'' (given these assumptions) self-consistent solutions to Eqs.~\ref{eqn:Ohms}-\ref{eqn:vdrift}, and we do so in \S~\ref{sec:extra.multi} for several examples. However this must be done numerically and is not intuitive nor applicable to all numerical implementations of non-ideal MHD. To gain insight and develop simple approximate fitting functions, we can use our three-species model in \S~\ref{sec:deriv}, and consider the asymptotic behavior in various limits. 

For $\epsilon \ll 1$ (the usual case), we have 
$\omega_{O} \rightarrow \omega_{-+} + \omega_{-n}$, 
$\omega_{H} \rightarrow \Omega_{2}$, 
$\omega_{A} \rightarrow \Omega_{3}\,\Omega_{4}/\omega_{+n}$, 
with 
$\Delta u_{-+}=|{\bf v}_{\rm drift}|=|{\bf v}_{\rm drift}^{0}|$ (independent of $\omega$), 
$\Delta u_{+n}=|{\bf v}_{\rm slip}^{+}| \rightarrow |{\bf v}_{\rm slip,\,\bot}|$, 
and 
$\Delta u_{-n} = |{\bf v}_{\rm slip}^{-}| \rightarrow v_{\rm drift,\,max}$. 
So $\omega_{H}\rightarrow \omega_{H}^{0}$. 
We then can solve Eq.~\ref{eqn:vdrift} exactly to obtain $\Delta u_{+n} \equiv v_{\rm slip}^{\rm eff} = v_{\rm T,\,eff}\,\{-1/2 + (1/4 +   |{\bf v}_{\rm slip,\,\bot}^{0}|^{2} / v_{\rm T,\,\rm eff}^{2} )^{1/2}\}^{1/2}$, and $\omega_{A} \rightarrow \omega_{A}^{0}\,[1/2 + (1/4 +  |{\bf v}_{\rm slip,\,\bot}^{0}|^{2} / v_{\rm T,\,\rm eff}^{2} ) ]^{-1/2} = \omega_{A}^{0} / \sqrt{1 + |v_{\rm slip}^{\rm eff}|^{2}/v_{\rm T,\,eff}^{2}}$. Then $\omega_{O} \rightarrow \omega^{0}_{-+}\,(1 + |{\bf v}_{\rm drift}^{0}|^{2}/v_{\rm T,\,eff,-}^{2})^{1/2} + \omega^{0}_{-n}\,(1 +  |{\bf v}_{\rm drift}^{0}|^{2}/v_{\rm T,\,eff,\,-}^{2} + |v_{\rm slip}^{\rm eff}|^{2}/v_{\rm T,\,eff,\,-}^{2})^{1/2}$. 
For $\epsilon \gg 1$ (e.g.\ all negative charge in dust grains), the same exercise gives: 
$\omega_{O} \rightarrow \omega_{-+} (1+\epsilon) + \omega_{+n} \rightarrow \omega_{-+}^{0} (1+\epsilon)\,[1 + 1 + |{\bf v}_{\rm drift}^{0}|^{2} / v_{\rm T,\,eff}^{2}]^{1/2} + \omega_{+n}^{0}\,[1 + |v_{\rm drift,\,max}|^{2} / v_{\rm T,\,eff}^{2}]^{1/2}$, 
$\omega_{H} \rightarrow -\Omega_{3} = \omega_{H}^{0}$, 
$\omega_{A} \rightarrow \Omega_{3}\,\Omega_{4}/\omega_{-n} \rightarrow \omega_{A}^{0} / \sqrt{1 + |v_{\rm slip}^{\rm eff}|^{2}/v_{\rm T,\,eff}^{2}}$, 
$\Delta u_{-n} \rightarrow  |{\bf v}_{\rm slip,\,\bot}| = v_{\rm slip}^{\rm eff} = v_{\rm T,\,eff}\,\{-1/2 + (1/4 +   |{\bf v}_{\rm slip,\,\bot}^{0}|^{2} / v_{\rm T,\,\rm eff}^{2} )^{1/2}\}^{1/2}$, 
$\Delta u_{+n} \rightarrow v_{\rm drift,\,max} = [1 +( | {\bf v}_{\rm drift}^{0}|^{2} + |v_{\rm slip}^{\rm eff}|^{2})/v_{\rm T,\,eff}^{2}]^{1/2}$. 
In both these limits, we can further note the expression for $\omega_{O}$ is quite closely approximated, for all salient limits of physical interest, with $\omega_{O} \approx \omega_{O}^{0}\,[1 + |v_{\rm drift,\,max}|^{2} / v_{\rm T,\,eff}^{2}]^{1/2}$ (there is no salient limit where the $\omega_{-+}$ term, which scales with $v_{\rm drift}$, is dominant in $\omega_{O}$ while also featuring  $\eta_{O} \gg \eta_{A,H}$ and $v_{\rm slip} \gg v_{\rm drift}$, which together defines the regime where this approximation might break down). 

This suggests the following simple approximation:
\begin{align}
\nonumber \eta^{\rm Eps}_{O} &\approx  \eta_{O}^{0}\,\left[ 1 +  \frac{v_{\rm slip,\,eff}^{2}+ |{\bf v}_{\rm drift}^{0}|^{2}}{v_{T,\,\rm eff}^{2}} \right]^{1/2} \ , \\ 
\nonumber \eta_{H}^{\rm Eps} &\approx \eta_{H}^{0} \ ,  \\ 
\label{eqn:epstein.scalings}\eta^{\rm Eps}_{A} &\approx \eta_{A}^{0} \,\left[ 1 +  \frac{v_{\rm slip,\,eff}^{2}}{v_{T,\,\rm eff}^{2}} \right]^{-1/2} \ .
\end{align}
Even in the ``worst case'' limit, with e.g.\ $\epsilon = 1$ and $\omega_{+n}=\omega_{-n}$ (e.g.\ all charge in identical positive/negative grains), these scalings are within a factor $\sim \sqrt{2}$ of the exact solution for these assumptions.

Intuitively, this scales as expected: $\eta_{O} \propto \omega$ increases with $\Delta u_{ij}$ in the superthermal limit, while $\eta_{H}$ is independent of $\omega$ to leading order, and $\eta_{A} \propto \omega^{-1}$ decreases with $\Delta u_{ij}$. The slip speed (as opposed to drift or slip+drift) appears in the scaling for $\omega_{A}$ because it is generally the slower/heavier (less-mobile) species whose relative velocity to the neutrals determines the effective ambipolar drift speed (Eq.~\ref{eqn:vdrift}), and this is modified to $v_{\rm slip,\,eff}$ as opposed to just $|{\bf v}_{\rm slip,\,\bot}^{0}|$ because the dependence of $v_{\rm slip}$ on $\eta_{A}$ (and therefore collision rates) in Eq.~\ref{eqn:vdrift} makes this an implicit nonlinear equation.\footnote{See also \citet{hillier:2024.ambipolar.drift.w.supersonic.vdrift}, who derive a similar correction for $\eta^{\rm Eps}_{A}$ in $v_{\rm slip,\,eff}$, but for the more restricted case of a two-fluid (hydrogen+proton) system with vanishing drift, neglecting Ohmic and Hall terms.} The total slip+drift speed appears in the Ohmic term because it is generally the faster/lighter (more mobile) species whose collisions with neutrals dominate (and in the $\omega_{-+}$-dominated regime, where $\Delta u_{-+}$ is just $v_{\rm drift}$, $v_{\rm slip}$ is generally small).

\subsection{Anomalous Terms}
\label{sec:anomalous.eta}

Now consider the anomalous term(s). Per \S~\ref{sec:superthermal:instabilities}, a rigorous derivation for the non-linear behaviors across scales of the wealth of different instabilities involved is not possible, but we can use our simple derivation from \S~\ref{sec:deriv} to gain a greater intuition and justification for the form of Eq.~\ref{eqn:etas.general}.

First, consider the better-studied case of anomalous resistivity from some instabilities or neglected terms from the charge carriers themselves. 
These are the examples discussed in \S~\ref{sec:superthermal:instabilities}, which manifest in Eq.~\ref{eqn:Ohms} or Eq.~\ref{eqn:induction.full} in different ways. Three specific examples are given there which we discuss in turn here. For case (1), we argued anomalous terms can appear in the Reynolds/bulk component of the pressure tensor (or battery) terms, giving rise to a terms like $\langle \nabla \cdot \mathbf{\Pi}_{-}\rangle \sim -(|q_{-}|\,n_{-}|\,|v_{\rm drift}|/\ell)\,{\bf v}_{\rm drift}$ in Eq.~\ref{eqn:Ohms}. Using the definition of ${\bf v}_{\rm drift}$, this can be rewritten as $ \sim \omega^{\rm an}\,|v_{\rm drift}/v_{T}|\,{\bf J}$. Inserting Eq.~\ref{eqn:Ohms} for ${\bf E}$ into the induction equation, we see this is equivalent to linearly adding a term to the Ohmic resistivity  
\begin{align}
\eta_{O} \rightarrow \eta_{O}^{0} + \eta_{O}^{\rm an}
\end{align}
with 
\begin{align}
\eta_{O}^{\rm an} \sim \frac{c^{2} m_{-}}{4\pi\,|q_{-}|^{2}\,n_{-}}\,\omega^{\rm an}\,\left| \frac{v_{\rm drift}}{v_{T}} \right|
\end{align}
 in the superthermal ($v_{\rm drift} \gg v_{T}$) limit. 
If we further assume the fluctuations/instabilities extend down to a scale $\ell$ given by the larger of either the gyro-radius or Debye length, then $\omega^{\rm an}$ scales as the gyro or plasma frequency.
For case (2), as noted in \S~\ref{sec:superthermal:instabilities} various studies have treated anomalous terms as arising from fluid-like but micro/meso-scale fluctuations, in a Reynolds-average sense, so they appear as terms like $\langle \delta {\bf U} \times \delta {\bf B} \rangle_{\mathcal{V}}$  in Eq.~\ref{eqn:induction.full}, with non-linear amplitude in the superthermal limit of $\sim |v_{\rm drift}|\,B\,\hat{\bf v}_{\rm drift}$. 
Given that this simply adds to the ``classical'' Ohmic term in the Reynolds-expanded induction Eq.~\ref{eqn:induction.full}, we see from this and Eq.~\ref{eqn:vdrift}-\ref{eqn:eta.units} (defining the units of $\eta_{O}$ and $v_{\rm drift}$) that this is again equivalent to linearly adding $\eta_{O} \rightarrow \eta_{O}^{0} + \eta_{O}^{\rm an}$, with the same scaling of $\eta^{\rm an}$ as above and $\omega^{\rm an} \sim \Omega_{1}$ of order the gyro frequency (for the plasma frequency-dominated high-density case, see references in \S~\ref{sec:superthermal:instabilities}). 
Case (3) described in \S~\ref{sec:superthermal:instabilities}, setting growth equal to damping of small-scale instabilities, treats the anomalous terms as giving rise to an additional effective (kinetic) collision rate between charge-carriers, i.e.\ $\omega^{\rm coll}_{-+} \rightarrow \omega^{\rm coll,\,0}_{-+} + \omega_{-+}^{\rm coll,\,an}$, where $\omega_{-+}^{\rm coll,\,an}$ is of order some fastest frequency (typically gyro or plasma) of the system times $|v_{\rm drift}|/v_{T}$. From the derivation in \S~\ref{sec:deriv}, we see that $\omega_{-+}^{\rm coll}$ appears only in $\omega_{O}$ linearly (through the $\omega_{1}$ term), and in no other terms arriving in Eq.~\ref{eqn:nonideal.classical}. Combining this with the fact that $v_{\rm drift}$ is set by ${\bf J}$ (independent of $\omega$), this is equivalent to linearly adding to $\eta_{O}$ as $\eta_{O} \rightarrow \eta_{O}^{0} + \eta_{O}^{\rm an}$ with the same form of $\eta_{O}^{\rm an}$ as above.

Thus, however we heuristically derive or argue for an anomalous resistivity, it appears with the same functional form, linearly adding to the classical $\omega_{O}^{0}$ with an effective scaling in the superthermal regime dimensionally given by the form we adopt in Eq.~\ref{eqn:etas.general}. 

There could, in principle, be anomalous terms that arise from superthermal slip as well. We ignore these here, for the reasons in \S~\ref{sec:superthermal:instabilities}: they are much less well-studied and their non-linear behavior is much more ambiguous. For example, because there is no assumption of confinement along magnetic fields for the neutral fluid, slip instabilities typically have linear and nonlinear Reynolds and stress terms that project along all of $\hat{\bf v}_{\rm drift}$, $\hat{\bf v}_{\rm slip,\,\bot}$,  $\hat{\bf v}_{\rm drift} \times \hat{\bf v}_{\rm slip,\,\bot}$ (i.e.\ all of the Ohmic, Hall, and ambipolar directions; \citealt{hopkins:2018.mhd.rdi,seligman:2018.mhd.rdi.sims,hopkins:2019.mhd.rdi.periodic.box.sims}), or even manifest as current-independent battery terms \citep{soliman:2024.dust.battery}. 
Moreover, many of the strongest superthermal slip instabilities depend on the external acceleration ${\bf a}_{\rm ext}$ terms in \S~\ref{sec:deriv} \citep[see][]{tytarenko:two.fluid.drift.instabilities,squire.hopkins:RDI}, but this introduces several new degrees of freedom and means both the saturation level and even directional sense of the associated terms depend on this unknown external ${\bf a}_{\rm ext}$. This also means that since the power source is ``external,'' the instabilities do not have to draw their energy from the currents ${\bf J}$, ${\bf J}_{d}$ (and therefore act resistively). 
In some cases slip instabilities could lead to stronger recoupling between charged and neutral fluids, acting like an anomalous charge-neutral collision term (which would suggest modifying $\eta_{A}$ as e.g.\ $\eta_{A}^{-1} \rightarrow (\eta_{A}^{0})^{-1} + (\eta_{A}^{\rm an})^{-1}$). But because there is no local restoring Coulomb force or equivalent of the local charge neutrality assumption (which acts to prevent total positive-negative charge separation), there is nothing to prevent slip instabilities from leading to total neutral-plasma separation (and indeed this is the saturation behavior seen in some MHD-PIC examples; \citealt{hopkins:2019.mhd.rdi.periodic.box.sims}), in which the system even on macro scales could require two-fluid treatments. Thus future work is needed to explore the super-thermal slip limit in more detail.
\,
\\ \,   
\\ \,   
\\ \,   

\end{appendix}

\end{document}